\documentclass[a4paper,11pt]{article}

\usepackage{jheppub}
\usepackage{xcolor,colortbl}
\allowdisplaybreaks

\usepackage{physics}
\usepackage{mathtools}
\usepackage{slashed}
\usepackage{subfigure}
\usepackage{dsfont}

\def\as{{\alpha_s}}

\def\pt{p_T}
\def\x{\xi}
\newcommand{\MathSpace}{\;\;\;\;\;\;\;\;\;\;\;\;}

\title{B-hadron hadro-production in NNLO QCD:\\ application to LHC $t\bar{t}$ events with leptonic decays}

\author[a]{Micha\l{} Czakon}
\author[a]{Terry Generet}
\author[b]{Alexander Mitov}
\author[b]{and Rene Poncelet}
\affiliation[a]{Institut f\"ur Theoretische Teilchenphysik und Kosmologie, RWTH Aachen University,\\ D-52056 Aachen, Germany}
\affiliation[b]{Cavendish Laboratory, University of Cambridge, Cambridge CB3 0HE, UK}
\emailAdd{mczakon@physik.rwth-aachen.de}
\emailAdd{terry.generet@rwth-aachen.de}
\emailAdd{adm74@cam.ac.uk}
\emailAdd{poncelet@hep.phy.cam.ac.uk}

\abstract{We calculate, for the first time, the NNLO QCD corrections to identified heavy hadron production at hadron colliders. The calculation is based on a flexible numeric framework which allows the calculation of any distribution of a single identified heavy hadron plus jets and non-QCD particles.  As a first application we provide NNLO QCD predictions for several differential distributions of $B$ hadrons in $t\bar t$ events at the LHC. Among others, these predictions are needed for the precise determination of the top quark mass. The extension of our results to other processes, like open or associated $B$ and charm production is straightforward. We also explore the prospects for extracting heavy flavor fragmentation functions from LHC data.}

\dedicated{\rm TTK-21-05, P3H-21-011, Cavendish-HEP-21/02}

\begin{document}
\maketitle
\flushbottom

\section{Introduction}

The production of heavy flavors, like bottom and charm, is a cornerstone high-energy collider process. It offers a wealth of information about the Standard Model and represents an excellent tool for probing the QCD dynamics. Heavy flavor production has been extensively studied at past and present high-energy lepton and/or hadron colliders as well as in nuclear collisions where heavy flavors are a prominent probe of the underlying nuclear dynamics. 

Heavy flavors are copiously produced at the LHC. Indeed, the $b\bar b$ and $c\bar c$ cross sections are among the largest at this collider. Such large production rates enable detailed and very precise measurements in wide kinematic ranges. The theoretical description of these processes, currently at next-to leading order (NLO) in QCD, is lagging in precision behind the experimental needs. For improving the precision of theory predictions the inclusion of the NNLO QCD corrections is mandatory.

When discussing the production of a heavy flavor of mass $m$ at a hadron collider, it is instructive to distinguish two kinematic regimes: the low $\pt$ regime where $\pt\sim m$ and the high $\pt$ one where $\pt\gg m$. The low $\pt$ production of a heavy flavor can be described in fixed order perturbation theory as an expansion in powers of the strong coupling constant evaluated at the scale $m$, i.e. $\as(m)$, and including the full dependence of the heavy quark mass $m$. For bottom, and especially charm, this expansion converges slowly since $\as(m)$ is not much smaller than unity. This expectation was confirmed by the recent fully-differential NNLO QCD calculation of $b\bar b$ production at the quark level \cite{Catani:2020kkl}. Such behavior is to be contrasted with $t\bar t$ production which is very similar technically but the smallness of $\as(m_t)$ leads to a well-converging perturbative expansion \cite{Czakon:2015owf,Czakon:2016dgf} through NNLO in QCD. 

The description of heavy flavor production at high $\pt$ involves a different set of challenges. Fixed order perturbation theory is no longer adequate there since large quasi-collinear logarithms $\log(\pt/m)$ appear to all orders in perturbation theory and need to be resummed. The resummation of these logs can be consistently carried out in the so-called perturbative fragmentation function (PFF) formalism \cite{Mele:1990cw}. Unlike the low $\pt$ case, a calculation of heavy flavor production at high $\pt$ is performed with a massless heavy quark since in the high-energy limit all terms that are power suppressed with $m$ are negligible while the mass-independent terms as well as the logarithmically enhanced ones are automatically accounted for by the PFF formalism. The current state of the art is NLO with next to leading logarithmic (NLL) accuracy. The goal of the present paper is to extend, for the first time, this description at hadron colliders to NNLO in QCD.

A generic application to heavy flavor production that is valid in all kinematic regimes would require the merging of the low $\pt$ and high $\pt$ descriptions mentioned above. This has been achieved at NLO in QCD within the so called FONLL approach \cite{Cacciari:1998it}. Some of the recent hadron collider applications include refs.~\cite{Cacciari:2012ny,Cacciari:2015fta}. Its generalization to NNLO goes beyond the scope of this paper. 

As a first application of this formalism at NNLO in QCD we compute several $B$-hadron differential distributions in top quark pair production and decay at the LHC $pp \to t\bar t+X\to B+X$. The reason for choosing this process is twofold: first, $B$-production is central to top quark physics and $B$-hadron related observables are a great tool for precise top quark mass determination at hadron colliders. Second, in $t\bar t$ events the top quark mass provides a natural large hard scale such that for almost all distributions of interest the power suppressed effects $\sim(m_b)^n$ are negligible. This makes this process an ideal application for the massless $b$ quark PFF formalism used in this work. $B$-hadron production in other processes, like open $B$ production at high $\pt$, would be a straightforward extension of the current work and we hope to report on it in future publications.

This work is organized as follows: in sec.~\ref{sec:general} we discuss the general features of the formalism for calculations with an identified hadron. In sec.~\ref{sec:computational-framework} we explain our calculational framework. In sec.~\ref{sec:PFF-NPFF} we introduce the $B$ fragmentation functions used in this work. Sec.~\ref{sec:applications} is devoted to phenomenological LHC applications. We study in detail $B$-hadron distributions in top quark decay and in $t\bar t$ production and decay. We also propose an observable which we find suitable for extracting $B$-hadron fragmentation functions from LHC data. Several appendices contain additional results. In appendix \ref{sec:tt-xsec-fragmentation} we give the structure of the NNLO cross section for the process $pp \to t\bar t+X\to B+X$. In appendix \ref{sec:collinear} we give in explicit form the general expressions for the collinear counterterms needed for any NNLO hadron collider process with fragmentation. Appendices \ref{sec:e+e-} and \ref{sec:sum-rules} present two highly non-trivial checks of our calculational setup: the calculation of $B$ production in $e^+e^-$ collisions which is compared to the exact analytic result and the fulfillment of sum rules in top quark decay.

\section{Fragmentation: the general framework}\label{sec:general}

A typical calculation in perturbative QCD involves final states with QCD partons, which are clustered into jets, and colourless particles such as leptons. By clustering particles into jets, information is lost about the properties of the individual particles. On the experimental side, it also introduces jet energy scale uncertainties, which can dominate the total uncertainty on jet-based observables (see e.g. ref.~\cite{Aad:2015nba}), but are largely absent when instead measuring a single hadron's momentum (e.g. ref.~\cite{Khachatryan:2016pek}). As an alternative to this usual approach of jet-based observables, it therefore seems appealing to instead consider observables involving the momentum of a single hadron, $h$.

Perturbation theory alone cannot describe non-perturbative phenomena like the transition from partons to hadrons, called fragmentation. The solution is to factorise the non-perturbative aspects into fragmentation functions \cite{Berman:1971xz} in analogy to how parton distribution functions are introduced to describe transitions from hadrons to partons in the initial state. The fragmentation functions depend on the hadron $h$ but are otherwise universal and can thus be extracted from experimental data. 

The theoretical description of the production of an identified hadron proceeds as follows. Standard tools and techniques are used to describe the production of on-shell partons. The partonic calculation is then extended by fragmenting the final-state partons, one at a time, into the observed hadron $h$ which has a well-defined momentum $p_h$. In practice, fragmentation corresponds to multiplying the fragmenting parton's momentum with a momentum fraction between $0$ and $1$, and then integrating the partonic cross section over it with a weight given by the corresponding fragmentation function. This procedure is equivalent to convolving the differential partonic cross sections with fragmentation functions:
\begin{equation}
\frac{d\sigma_h}{dE_h}(E_h) =  \sum_i \bigg(D_{i\to h}\otimes \frac{d\sigma_i}{dE_i}\bigg)(E_h) \equiv \sum_i \int_{0}^1\frac{dx}{x} D_{i\to h}(x)\frac{d\sigma_i}{dE_i}\bigg(\frac{E_h}{x}\bigg)\;,
\label{eq:factorized-x-section}
\end{equation}
where the summation over $i$ is over all partons in the final state. $D_{i\to h}$ is the fragmentation function for the transition $i\to h$. Although the hadron's energy $E_h$ is used as an example here, any observable linear in the hadron's momentum can be utilized. 

The kinematics of the collinear fragmentation process can be represented as follows
\begin{equation}
i(p_i) \to h(p_h) + X(p_i-p_h)\;,\;\;\;\; p_h^\mu= xp_i^\mu\;,\;\;\;\; x\in[0,1]\;,
\label{eq:splitting-kinematics}
\end{equation}
where the momenta of particles have been indicated in brackets and $X$ represents the particles produced in the fragmentation process which are not explicitly described by the fragmentation function, i.e. all particles in the jet initiated by $i$ other than the observed hadron $h$. Essentially, this means that one relates the hadron's momentum to that of a single parton, the latter being an infrared-unsafe quantity. 

As the above discussion indicates, the partonic cross section for producing a parton $i$ is infrared unsafe and therefore contains uncancelled divergences. These are collinear divergences which factorise into lower-order contributions to the cross section and process-independent splitting functions. Because of this general and process-independent structure, it is possible to absorb the uncancelled divergences into the fragmentation functions via collinear renormalisation \cite{Ellis:1991qj}:
\begin{equation}
D_{i\to h}^{\text{bare}}(x) = \sum_j \big(\hat\Gamma_{ij}\otimes D_{i\to h}\big)(x)\;,
\label{eq:collinear-ren}
\end{equation}
where the sum is over all partons. The collinear counterterms $\hat\Gamma_{ij}$ are functions of $x$ and can be specified, not uniquely, within perturbation theory. In practice a choice is made about the finite terms contained in these counterterms. Such a choice implies that the IR renormalized coefficient and fragmentation functions, $d\sigma_i$ and $D_{i\to h}$, are individually scheme dependent however their convolution $d\sigma_h$ is not, as one may expect from an observable. As for parton distribution functions, it is standard practice to define the counterterms $\hat\Gamma_{ij}$ in the $\overline{\rm MS}$ scheme.

The collinear renormalisation eq.~(\ref{eq:collinear-ren}) introduces scale dependence into the renormalised fragmentation functions, which is described by the (time-like) Dokshitzer-Gribov-Lipatov-Altarelli-Parisi (DGLAP) evolution equations \cite{Altarelli:1977zs,Dokshitzer:1977sg,Gribov:1972ri}:
\begin{equation}
\mu_{Fr}^2\frac{dD_{i\to h}}{d\mu_{Fr}^2}(x,\mu_{Fr}) = \sum_j \big(P^{\text{T}}_{ij}\otimes D_{j\to h}\big)(x,\mu_{Fr})\;,
\label{eq:DGLAP}
\end{equation}
where $P^{\text{T}}_{ij}$ are the time-like splitting functions, known through NNLO \cite{Mitov:2006ic,Moch:2007tx,Almasy:2011eq}, and $\mu_{Fr}$ is the fragmentation factorisation scale, or simply the fragmentation scale. Because fragmentation functions are extracted from experiment at a certain scale, it is necessary to relate fragmentation functions evaluated at two different scales. This is achieved by solving the DGLAP equations eq.~(\ref{eq:DGLAP}). The initial conditions necessary for fully specifying the solution are discussed in sec.~\ref{sec:PFF-NPFF}. The solution of the DGLAP equation has the additional benefit that any large logarithms of the ratio of two scales are resummed with a logarithmic accuracy given by the order of the splitting functions used.

\section{Computational approach}\label{sec:computational-framework}

Fixed order calculations are typically performed using a subtraction scheme. The purpose of a subtraction scheme is to ensure that in any singular limit of the kinematics, the singularities of physical cross sections are matched by those of the relevant subtraction terms and that in those limits, the corresponding final states are indistinguishable. These are the requirements for the numerical integrability of the cross section. If the singular behaviour of the cross section is not matched by its subtraction terms, then a numerically non-integrable singularity remains. If the singular behaviours of the contributions match, but the kinematics are distinct, then the fully inclusive cross section is numerically integrable, but differential and fiducial cross sections may not be. Schematically, a cross section differential in some observable $O$ can be written as
\begin{align}
\frac{d\sigma}{dO} &= \sum_n\int_n d\sigma_n(\{y_i\}_n)\delta(O(\{y_i\}_n)-O) \notag\\
&= \sum_n\int_n \bigg(d\sigma_n\delta(O(\{y_i\}_n)-O)-\sum_m d\sigma_n^m\delta(O^m(\{y_i\}_n)-O)\notag\\
&\phantom{{}= \sum_n\int_n \bigg(d\sigma_n\delta(O(\{y_i\}_n)-O)}+\sum_m d\sigma_n^m\delta(O^m(\{y_i\}_n)-O)\bigg)\;,
\label{eqn:SubtScheme}
\end{align}
where $n$ denotes the number of final-state particles, $d\sigma_n$ is the fully differential $n$-particle cross section, $\{y_i\}_n$ is the set of $n$-particle phase space parameters to be integrated over, e.g.\ the set of momentum components of the particles, $d\sigma_n^m$ is a subtraction term, the integral $\int_n$ is over the full $n$-particle phase space and the dependence of $d\sigma_n$ and $d\sigma_n^m$ on $\{y_i\}_n$ has been omitted on the second and third lines for brevity. As usual, the intent is to integrate the combination of terms on the second line fully numerically, while the integration over the singular behaviour of the terms on the third line is performed analytically.

Due to conceptual differences, a subtraction scheme has to be modified with respect to the case without fragmentation in order to perform calculations involving fragmentation. Such modifications have been made to the sector-improved residue subtraction scheme \cite{Czakon:2010td,Czakon:2011ve,Czakon:2014oma,Czakon:2019tmo} and its implementation in the {\tt Stripper} library, enabling the calculations presented here. The additional complications due to fragmentation have been discussed in the past in the context of NLO subtraction schemes, see e.g. ref.~\cite{Catani:1996vz}, and no further complications are introduced beyond NLO. Nonetheless, the required modifications will also be discussed here for consistency and completeness. All of the necessary changes can be identified by considering which additional requirements fragmentation effectively puts on the calculation. Writing the fragmentation equivalent of eq.\ \eqref{eqn:SubtScheme}, these requirements become apparent:
\begin{align}
\frac{d\sigma}{dO} &= \sum_n\int_0^1dx\int_n d\sigma_n(\{y_i\}_n)D(x)\delta(O(\{y_i\}_n,x)-O) \notag\\
&= \sum_n\int_0^1dx\int_n \bigg(d\sigma_n(\{y_i\}_n)D(x)\delta(O(\{y_i\}_n,x)-O)\notag\\
&\phantom{{}= \sum_n\int_0^1dx\int_n \bigg(}-\sum_m d\sigma_n^m(\{y_i\}_n)D(\tilde{x}^m(\{y_i\}_n,x))\delta(O^m(\{y_i\}_n,x)-O)\notag\\
&\phantom{{}= \sum_n\int_0^1dx\int_n \bigg(}+\sum_m d\sigma_n^m(\{y_i\}_n)D(\tilde{x}^m(\{y_i\}_n,x))\delta(O^m(\{y_i\}_n,x)-O)\bigg)\;,
\label{eqn:SubtSchemeFrag}
\end{align}
where for simplicity a single fragmentation function contributes. A realistic cross section would simply be a sum over such contributions. The functions $\tilde{x}^m$ will be discussed later. The two differences with respect to eq.\ \eqref{eqn:SubtScheme} are multiplication by the fragmentation function and the dependence of the observable on the momentum fraction $x$. For subtraction terms, the dependence of the observable on the phase space parameters changes as well and this point will be discussed first.

\begin{figure}[t]
	\centering
	\includegraphics[width=1\textwidth]{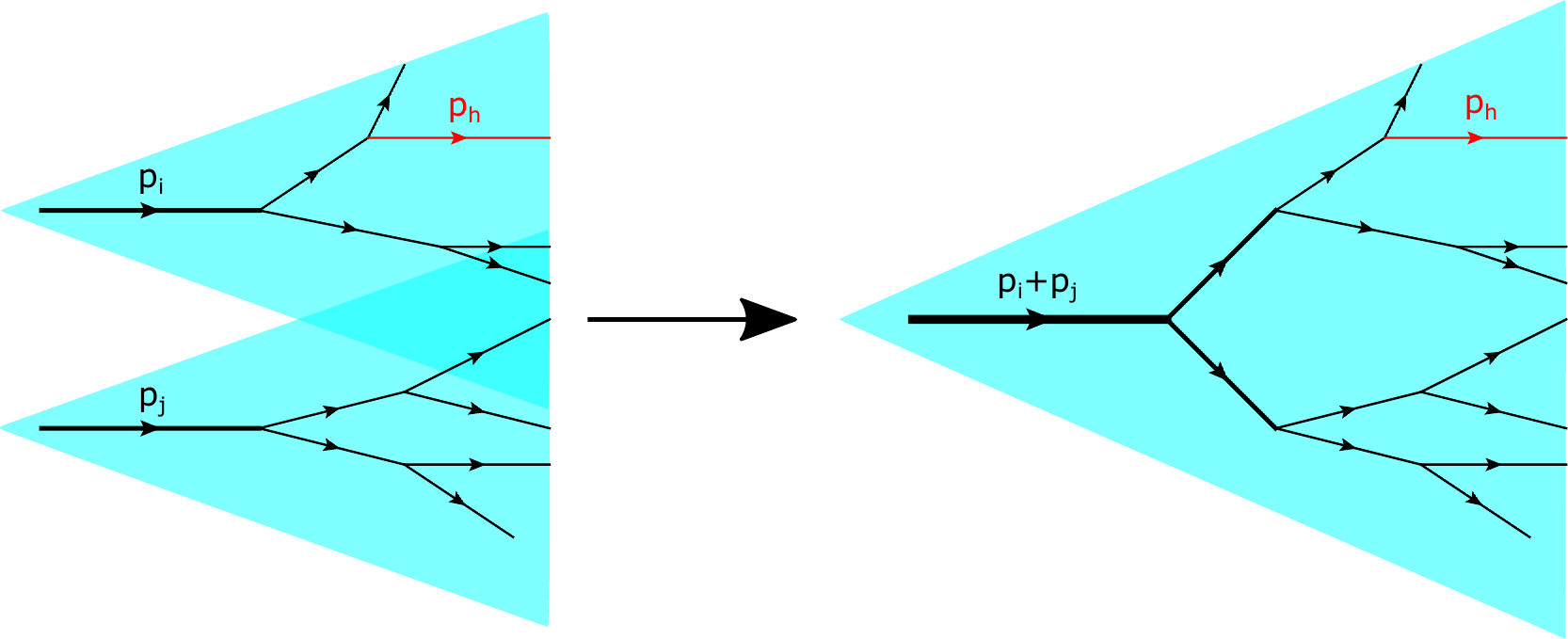}
	\caption{Observable kinematics in a collinear limit. When the partons $i$ and $j$ become collinear, the jets they initiate become indistinguishable from a jet initiated by a single particle carrying the sum of their momenta (blue shaded regions). If a single hadron $h$ is identified in the final state (red), then the fraction of the fragmenting particle's momentum carried by $h$ is smaller for the combination of $i$ and $j$ than for $i$, since the momentum $p_h$ does not change.}
	\label{fig:CollLimit}
\end{figure}

In typical calculations without fragmentation, all partons are clustered into jets and all observables depend only on the kinematics of the partons indirectly through the kinematics of jets. For a collinear limit, this means the relative magnitude of the momenta of the collinear partons is irrelevant, as only their sum enters the observable. Because of this, it is sufficient for the kinematics of the subtraction term to correspond to the exact collinear configuration, replacing the collinear partons by a single parton carrying the appropriate combination of their conserved quantities, such as flavour and momentum, as illustrated in fig.\ \ref{fig:CollLimit}. If one of the collinear partons fragments, then the magnitude of its momentum does enter observables, as it is directly related to the momentum of the final hadron via the rescaling by the momentum fraction. For the example shown in fig.\ \ref{fig:CollLimit}, this implies the requirement
\begin{equation}
x_i p_i = p_h = x_{ij} (p_i + p_j)\;,
\end{equation}
where $x_i$ and $x_{ij}$ are the momentum fractions for the parton $i$ and the combination of $i$ and $j$, respectively. Similarly, the flavour of the fragmenting parton determines the size of the contribution through the fragmentation function. When moving to the subtraction kinematics, it is thus necessary to retain the information on the contribution from the fragmenting parton to the total momentum of the collinear partons, which for the example above is e.g.\ the ratio between $p_i$ and $p_i+p_j$, and the flavour of the fragmenting particle.

There is an important point to stress here concerning soft limits, as the situation is slightly different. A singular soft limit occurs when the total energy of a flavourless set of partons -- containing gluons and equal numbers of quarks and anti-quarks of each flavour -- becomes small. The standard observation is that a configuration containing a zero-energy flavourless set of partons cannot be distinguished from one where this set is removed from the final state. The kinematics of the subtraction term thus corresponds to the exact soft configuration, removing the zero-energy flavourless set of partons from the final state. The statement that zero-energy, flavourless sets of partons can be removed from a final state without changing any observable is no longer true if one of those partons fragments, as the hadron is assumed to always be observable on its own. One could in principle proceed as for the collinear case and construct the subtraction kinematics as usual, keeping the information about the flavour and momentum, the latter being zero by definition, of the fragmenting parton in the soft limit. However, this yields contributions where the hadron always carries zero momentum. Not only is this an unphysical configuration, as the hadron has a non-zero mass, the factorisation of the cross section into the hard process and a fragmentation function only applies if the hard scale of the hadron, e.g.\ its transverse momentum, is much larger than its mass \cite{Collins:1989gx}. Additionally, fragmentation functions are divergent as the momentum fraction goes to zero, so even if these soft limits of the partonic cross section were regulated, the hadronic cross section would still be divergent. Because of this, there are no (integrated) subtraction terms regulating the soft limit of a fragmenting parton.

By considering exact singular limits, it has been explained that the kinematics of subtraction terms must be modified. The exact dependence of the kinematics on the full phase space parametrisation is arbitrary, only in singular limits must the kinematics of the cross section and its corresponding subtraction terms match, i.e.:
\begin{equation}
O^m(\{y_i\}_n,x) \to O(\{y_i\}_n,x)\;,
\label{eqn:ObservableLimit}
\end{equation}
where the limit is any limit which is supposed to be regulated by $d\sigma_n^m$ and eq.\ \eqref{eqn:ObservableLimit} should hold for all infrared-safe observables, where the momentum of the hadron is considered an infrared-safe quantity within this framework. Without fragmentation, the analytic integration performed to obtain the integrated subtraction term relies on the fact that $O^m$ does not depend on the parameters integrated over. As explained above, this may not be the case when one of the partons fragments. An example would be a subtraction term which regulates both a collinear and a soft singularity, which is a part of e.g.\ the Catani-Seymour dipole subtraction scheme \cite{Catani:1996vz}, with one of the partons fragmenting. In this case, the energy of the hadron would depend on the energy of the soft parton, so it would depend on the parameter parameterising the soft limit, spoiling the ability to perform this integration fully analytically. An implementation of a subtraction scheme containing such subtraction terms would therefore require laborious modifications before general fragmentation computations can be performed.

Here a critical simplification exists for the sector-improved residue subtraction scheme with respect to many other subtraction schemes. If there are no subtraction terms which regulate more than one type of singularity, i.e.\ every subtraction term is designed to counter a singularity occurring as all elements of a single set of phase space parameters simultaneously approach a singular point, then the kinematics of each subtraction term can be chosen to always match those of the cross section in the subtraction term's characteristic singular limit. Because this is a constant with respect to the variables which parameterise the subtraction term and are integrated over analytically in the integrated subtraction term, the integrated subtraction is unchanged by the introduction of fragmentation (aside from an overall factor given by the fragmentation function), avoiding the need to redo any analytic integration previously performed for the original subtraction scheme. For this reason, the sector-improved residue subtraction scheme is particularly suited for the extension to fragmentation, since it does not contain any subtraction terms which regulate multiple singularities \cite{Czakon:2014oma}.

Aside from changes to the kinematics of the final state, the inclusion of fragmentation also modifies the size of the contributions of different phase space points via multiplication by the fragmentation function, as shown in eq.\ \eqref{eqn:SubtSchemeFrag}. There is a certain amount of freedom when it comes to the point at which the fragmentation function is evaluated for a subtraction term, written in eq.\ \eqref{eqn:SubtSchemeFrag} as the functions $\tilde{x}^m(\{y_i\}_n,x)$. The only strictly necessary condition is that a subtraction term matches the singular behaviour of the cross section in certain singular limits. This requires that in any singular limit, the fragmentation function is evaluated at the same point for both the cross section and its corresponding subtraction terms, i.e.:
\begin{equation}
\tilde{x}^m(\{y_i\}_n,x) \to x\;,
\end{equation}
where the limit is again any limit which is supposed to be regulated by $d\sigma_n^m$. The most simple choice
\begin{equation}
\tilde{x}^m(\{y_i\}_n,x) \equiv x
\label{eqn:SubtMomentumFraction}
\end{equation}
is made here. Note that in order to reuse the integrated subtraction terms from the case without fragmentation as explained above, $\tilde{x}^m(\{y_i\}_n,x)$ must fulfill an additional condition: it should not depend on the parameters parameterizing the singular limit regulated by $d\sigma_n^m$. This is trivially satisfied by the choice shown in eq.\ \eqref{eqn:SubtMomentumFraction}.

The modifications discussed up until now are sufficient to perform calculations with fragmentation, but often lead to suboptimal numerical convergence. The reason for this is that while the kinematics of the cross section and one of its subtraction terms match in the singular limit, they do not in the remainder of the phase space. It is thus possible for both contributions to be large with opposite signs, but instead of mostly cancelling each other, the contributions are added to different bins of a calculated histogram. This missed-binning increases the fluctuations within individual bins, increasing their Monte Carlo uncertainty for a given number of events and thus reducing the rate of numerical convergence. To mitigate this, one can rescale the momentum fraction $x$ for each contribution on an event-by-event basis, such that the value of an observable of choice is always identical for all contributions for any given event. If this reference observable is now binned in a histogram, then missed-binning cannot occur by definition, potentially vastly improving the numerical convergence.

The final difference with respect to calculations without fragmentation is the introduction of collinear renormalisation counterterms for fragmentation functions. These are conceptually identical to those for PDFs and it is well-known how to obtain them in terms of splitting functions. The only difference with respect to the renormalisation of PDFs is the need to use time-like splitting functions, which differ from the space-like ones starting at NLO \cite{Curci:1980uw,Furmanski:1980cm}. For completeness, in appendix \ref{sec:collinear} we present the explicit expressions for the collinear counterterms while in appendix \ref{sec:tt-xsec-fragmentation} we give in some detail the structure of the cross section for the process $pp\to t\bar t+X\to B+X$.

\section{Perturbative and Non-Perturbative Fragmentation Functions for Heavy Flavor Fragmentation}\label{sec:PFF-NPFF}

The fragmentation functions used in this paper are based on the perturbative fragmentation function approach \cite{Mele:1990cw}, in which all fragmentation functions for the production of heavy-flavoured hadrons can be related to a single non-perturbative fragmentation function (NPFF) via convolutions with perturbatively calculable coefficients, called perturbative fragmentation functions (PFFs):
\begin{equation}
D_{i\to h}(x) = \big(D_{i\to q}\otimes D_h^{\text{NP}}\big)(x)\;,
\end{equation}
where $i$ can be any parton, $h$ is the heavy-flavoured hadron and $q$ is the heavy quark. The heavy-quark PFFs were originally derived at NLO \cite{Mele:1990cw} and have since been computed at NNLO as well \cite{Melnikov:2004bm,Mitov:2004du}. The only ingredient required to compute FFs for the production of heavy-flavoured hadrons is thus the NPFF. Typically, NPFFs are extracted from $e^+e^-$ data, however, theoretically motivated ones also exist \cite{Braaten:1994bz,Aglietti:2006yf}.

In the remainder of this work we will be interested in the case where the heavy quark is the bottom, i.e. $q=b$, and the heavy-flavored hadron is a $b$-flavored one, i.e. $h=B$.

At present, no such extraction at NNLO employing the PFF approach is available in the literature. For this reason, three different sets of FFs were obtained from two different extractions, each set corresponding to a different compromise. A third extraction, which follows an approximation of the PFF approach, was presented in ref.~\cite{Salajegheh:2019ach}, but has not been used here.

The first two sets of FFs are based on the extraction of ref.~\cite{Fickinger:2016rfd}. The FF of that paper is not based on the PFF approach, instead relying on effective field theory calculations. Nonetheless, a NPFF was extracted at NLO and NNLO, including NNLL and N$^3$LL large-$x$ resummation, respectively. Unfortunately, due to the different approach to the computation of FFs, there is no simple relation between the FF of that paper and one computed within the PFF approach. A reasonable conversion from one type of FF to the other has to be chosen. Another important point is that the extracted FF corresponds to the non-singlet (NS) combination, i.e.\ the difference between the bottom and anti-bottom FFs:
\begin{equation}
D_B^{\text{NS}}(\mu_{Fr},x) = D_{b\to B}(\mu_{Fr},x)-D_{\bar{b}\to B}(\mu_{Fr},x)\;.
\end{equation}

The set of FFs used most centrally in this paper is labelled ``FFKM". Its initial conditions are obtained by taking the extracted non-singlet function of ref.~\cite{Fickinger:2016rfd} evaluated at the initial scale $\mu_{Fr}=\mu_0$ with $\mu_{0}=m_b=4.66$ GeV, then calculating the FFs other than the bottom-quark FF from the PFFs and the extracted NPFF and, finally, adding the anti-bottom FF to the non-singlet one to obtain the full bottom FF:
\begin{align}
D_{i\to B}(\mu_{0},x) &= \big(D_{i\to b}\otimes D_B^{\text{NP}}\big)(\mu_{0},x)\;,\;\;i\neq b\;,\\
D_{b\to B}(\mu_{0},x) &= D_B^{\text{NS}}(\mu_{0},x)+D_{\bar{b}\to B}(\mu_{0},x)\;.
\end{align}
The FFs at any other scale $\mu_{Fr} > \mu_0$ are then obtained by evolving these initial conditions using the DGLAP evolution library {\tt APFEL} \cite{Bertone:2013vaa}.

An alternative construction labelled ``FFKM(2)" is to proceed as for the FFKM set, but as a final step the non-singlet contribution at each scale is replaced by the non-singlet contribution at that scale as provided by the authors of ref.~\cite{Fickinger:2016rfd}. This is not equivalent to the FFKM set, since the FF of ref.~\cite{Fickinger:2016rfd} does not satisfy the non-singlet DGLAP evolution equation.

The third and final set of FFs, labelled ``CNO", is obtained by taking the extraction of ref.~\cite{Cacciari:2006vy}. This extraction was performed using the PFF approach, but only at NLO including NLL large-$x$ resummation. This time $\mu_{0}=m_b=4.75$ GeV.
\begin{table}
	\centering
	\begin{tabular}{| c || c | c | c | c |}
		\hline
		FF set & NPFF & PFF & Large-$x$ & DGLAP\\\hline\hline
		FFKM/FFKM(2) @ NLO & NLO & NLO & NNLL & NLL\\\hline
		FFKM/FFKM(2) @ NNLO & NNLO & NNLO & N$^3$LL & NNLL\\\hline
		CNO @ NLO & NLO & NLO & NLL & NLL\\\hline
		CNO @ NNLO & NLO & NNLO & NLL & NNLL\\\hline
	\end{tabular}
	\caption{The differences between the FFKM/FFKM(2) sets and the CNO set at NLO and NNLO in terms of perturbative and logarithmic orders. NPFF refers to the perturbative order at which the extraction of non-perturbative parameters was performed. PFF refers to the perturbative order of the PFFs used. The column ``large-$x$" shows the logarithmic order of the resummation of logarithms of $1-x$, while the column labelled ``DGLAP" indicates the logarithmic order of the DGLAP resummation.}
	\label{tab:FragFuncOrders}
\end{table}

NLO and NNLO versions of all three sets were constructed. The perturbative and logarithmic orders of different components of the fragmentation functions are shown in \linebreak table\ \ref{tab:FragFuncOrders}. All FFs are symmetrised with respect to particles and anti-particles. The scale evolution is always performed using {\tt APFEL}, where the value and running of $\alpha_s$ are always chosen to match those of the PDF set used at the same order. As an alternative to performing the evolution with {\tt APFEL}, the {\tt MELA} \cite{Bertone:2015cwa} library could have been used instead, as was e.g.\ done in ref.~\cite{Ridolfi:2019bch} to perform a detailed study of the evolution of heavy-quark fragmentation functions. For simplicity, neither {\tt MELA} nor the results of ref.~\cite{Ridolfi:2019bch} have been used here.

In order to be able to estimate uncertainties due to the errors on the extracted FFs, multiple versions of all sets were constructed, corresponding to taking the extracted non-perturbative parameters and independently varying them by one standard deviation. Since there is only one parameter for the FFKM and FFKM(2) sets, this leads to three variations each, while the CNO set involves two parameters, leading to 9 variations. For the CNO set, correlations between the parameters are ignored.

All three FFs were found to be within reasonable agreement with each other, suggesting none of the individual compromises are particularly significant.

\section{Applications}\label{sec:applications}

\subsection{$b$-fragmentation in top-quark decay}\label{sec:top-decay}

As a first application we consider the process $t\to B+W+X$ with the subsequent decay $W\to \ell+\nu$ in NNLO in QCD. We work with top quark pole mass $m_t = 172.5\;\text{GeV}$. We use fixed scale choices for the renormalization and fragmentation scales: $\mu_R=\mu_{Fr}=m_t/2$. The rationale for this scale choice is discussed in the next section. Scale variation is done following the standard 7-point scale variation approach: $1/2 \le \mu_{R}/\mu_{Fr} \le 2$. Perturbative calculations for top decay at any accuracy (LO, NLO or NNLO) are always convolved with FF at NNLO. In all cases the value of the strong coupling $\as$ is taken from the {\tt LHAPDF} interface \cite{Buckley:2014ana} as produced by the {\tt NNPDF3.1} NNLO pdf set \cite{Ball:2017nwa}. Further details about this process and its setup can be found in appendix \ref{sec:tt-xsec-fragmentation} as well as in ref.~\cite{Czakon:2020qbd}.

In all observables discussed in this section we implement an energy cutoff of $E(B) > 5$ GeV. This cutoff helps us avoid the low $x$ region of the FFs. Excluding this region is not consequential for this work since in our implementation all power corrections $\sim (m_b)^n, n\ge 2$, are neglected and our predictions are not valid in the very low $x$ region anyway.

As a check on our implementation we have verified that our calculation satisfies the momentum conservation sum rule, see appendix \ref{sec:sum-rules} for details.

We study the following observables: the invariant mass of the lepton and the hadron $m(B\ell)$ and the energy fraction of the $B$-hadron to its maximum energy $E(B)/E(B)_{\text{max}}$, where
\begin{equation}
 E(B)_{\text{max}} = \frac{m_t^2-m_W^2}{2 m_t}\,.
 \label{eq:EB-max}
\end{equation}

\begin{figure}[t]
  \centering
   \includegraphics[width=0.49\textwidth]{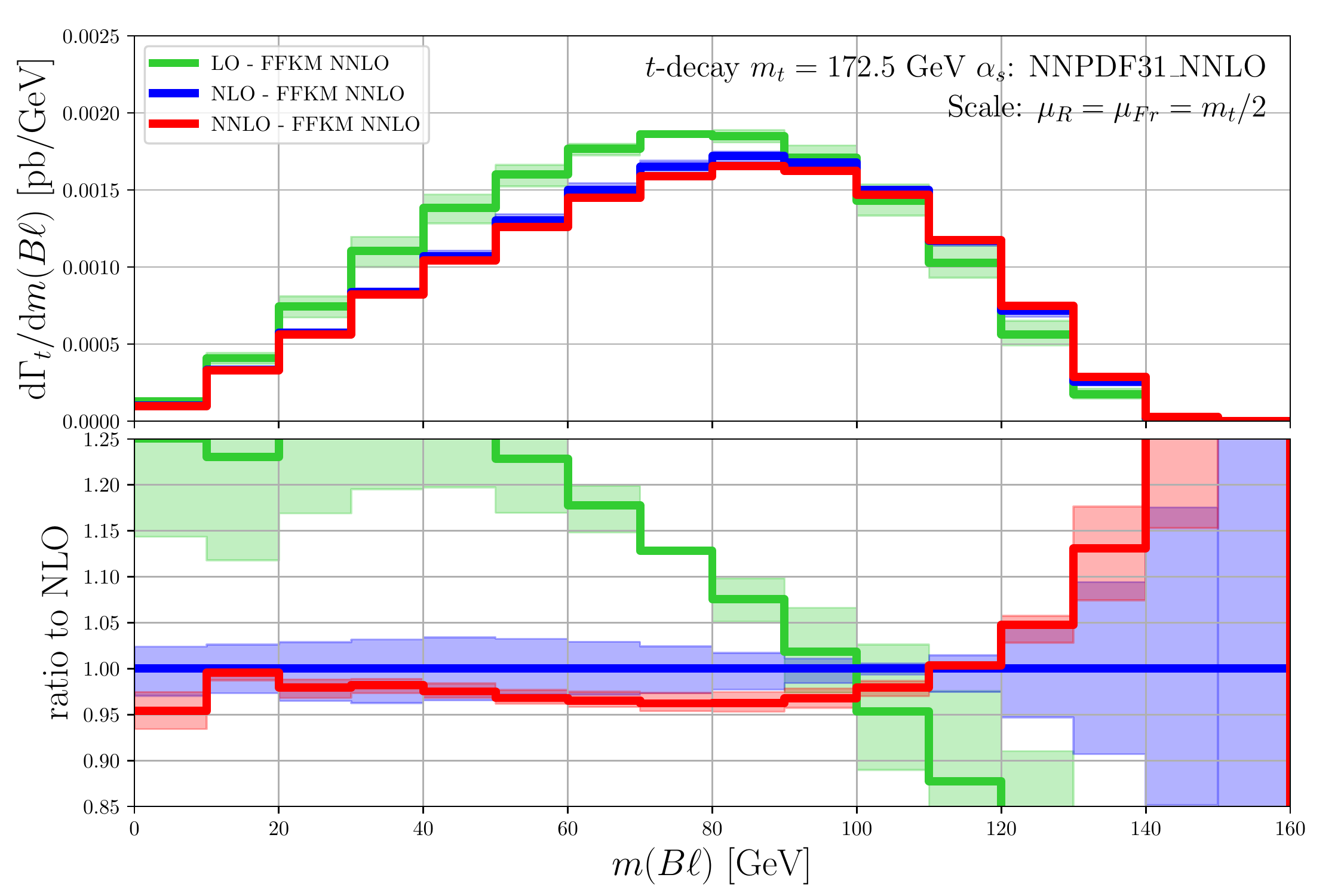}
   \includegraphics[width=0.49\textwidth]{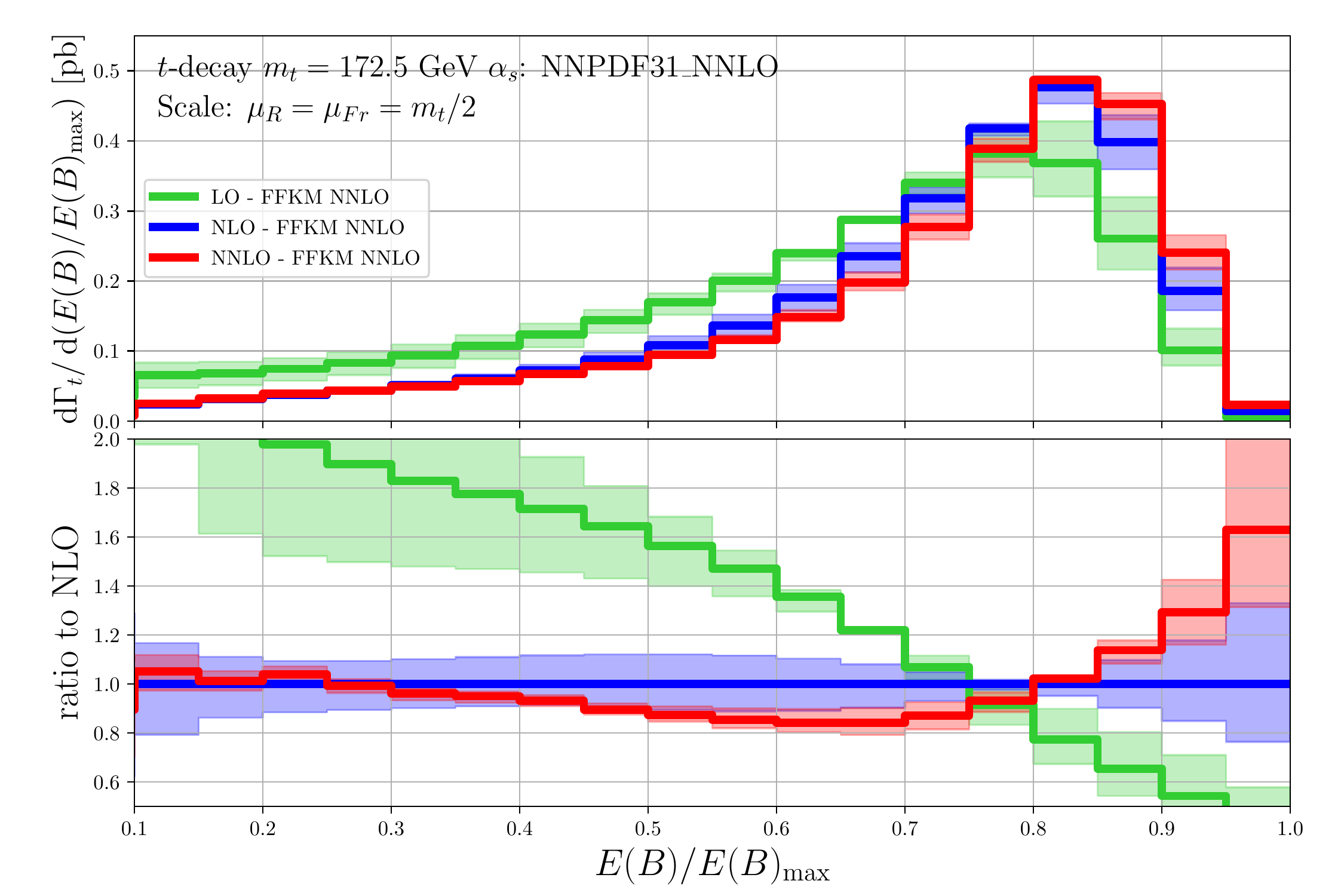}
  \caption{Absolute differential top decay width as a function of the invariant mass $m(B\ell)$ (left) and the energy fraction $E(B)/E(B)_{\text{max}}$ (right). All curves are convoluted with the same FF: FFKM at NNLO. Shown is comparison for different perturbative orders: LO, NLO and NNLO.}
  \label{fig:top-decay-order}
\end{figure}

\begin{figure}[t]
  \centering
   \includegraphics[width=0.49\textwidth]{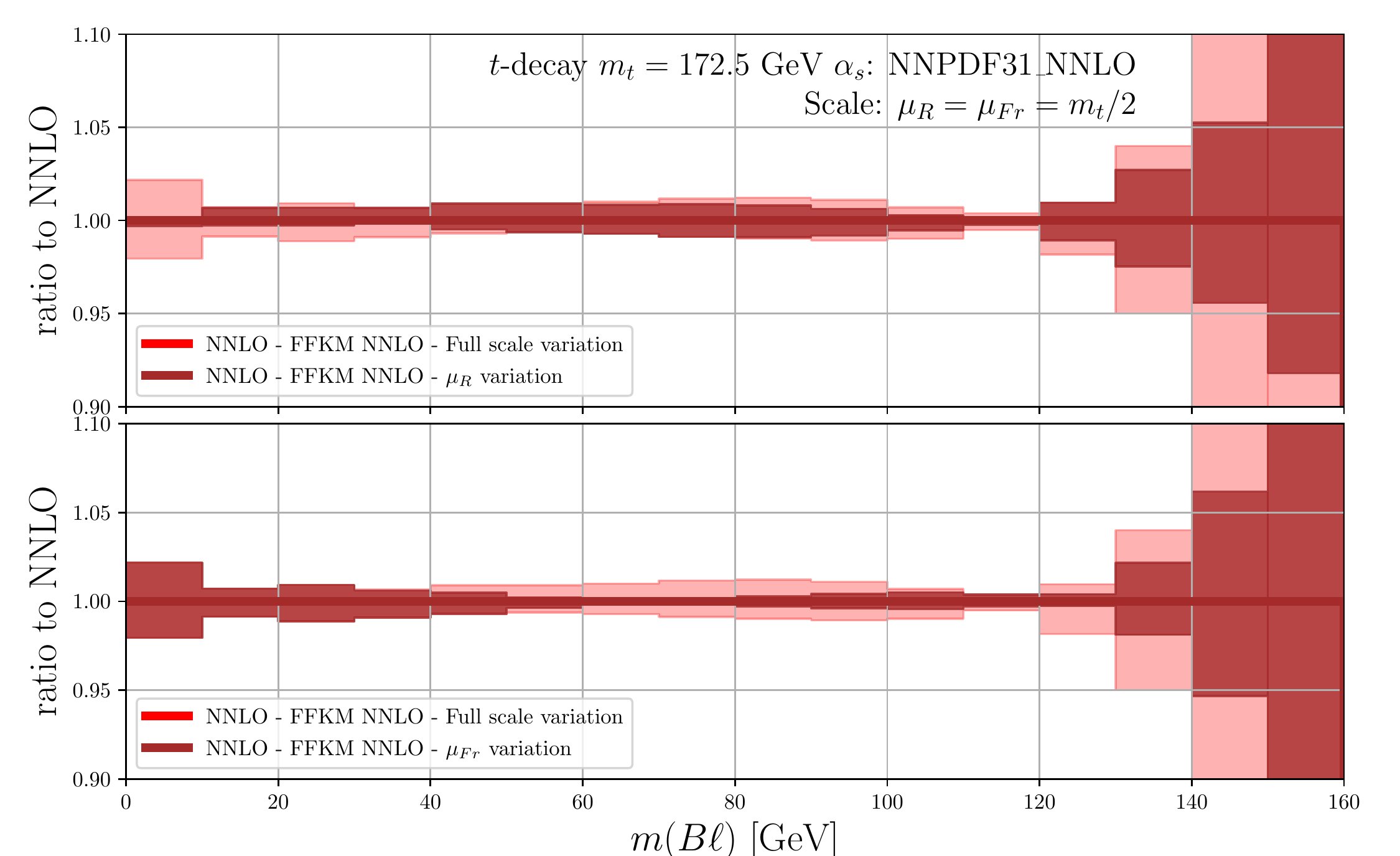}
   \includegraphics[width=0.49\textwidth]{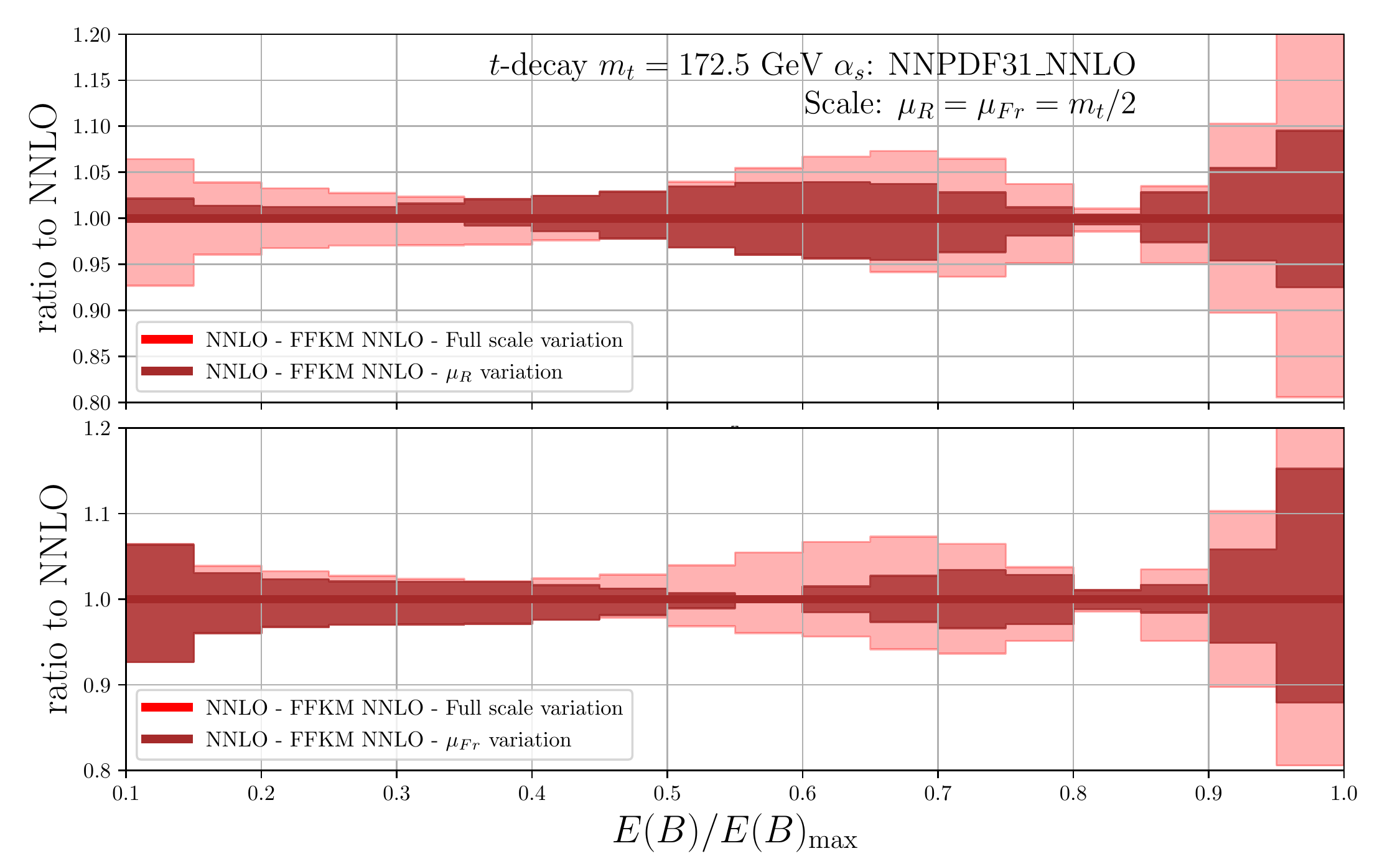}
  \caption{As in fig.~\ref{fig:top-decay-order} but showing the scale variation of the NNLO prediction: $\mu_R$-only vs. total scale variation (upper plot) and $\mu_{Fr}$-only vs. total scale variation (lower plot).}
  \label{fig:top-decay-scale}
\end{figure}

\begin{figure}[t]
  \centering
   \includegraphics[width=0.49\textwidth]{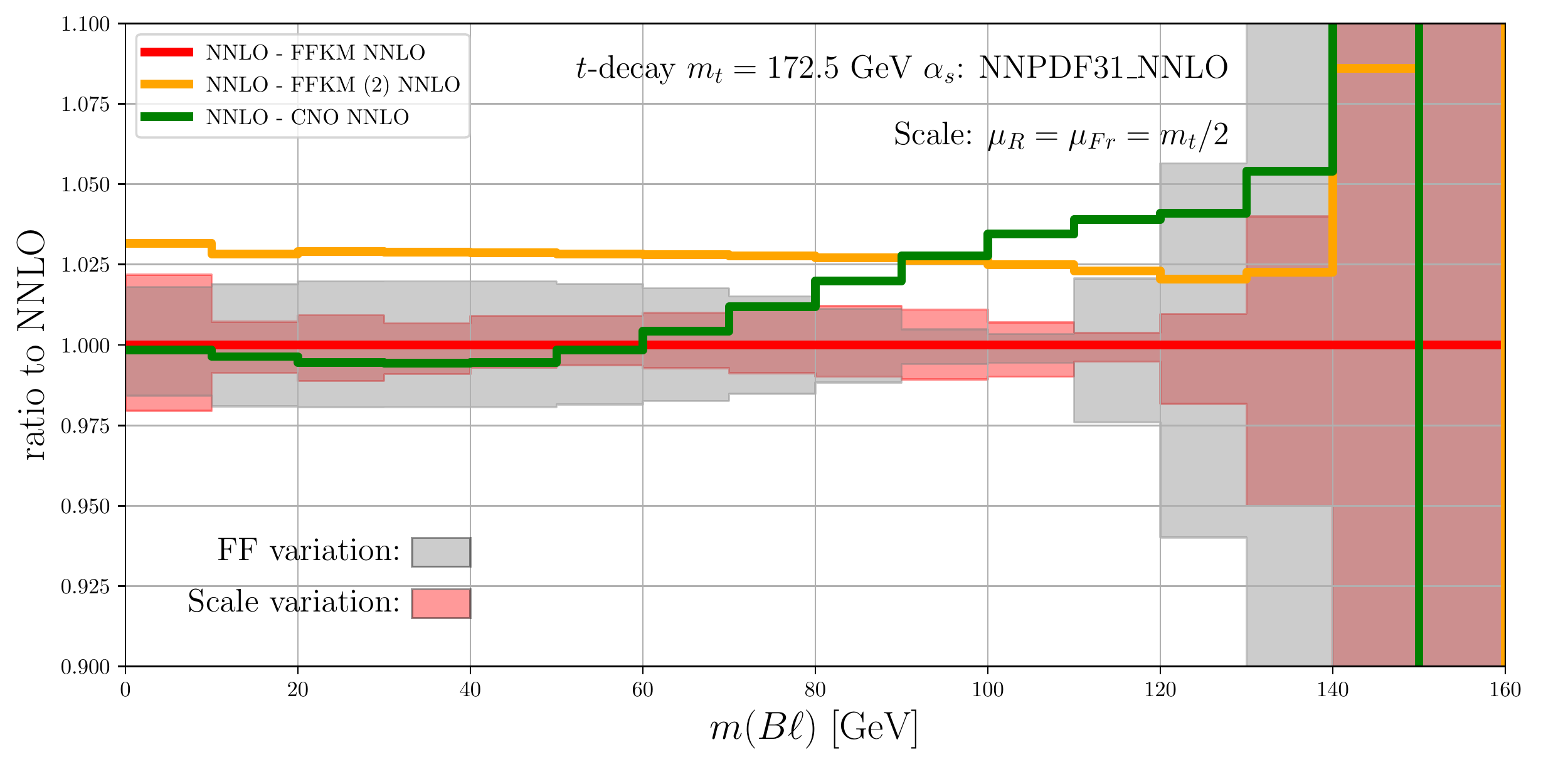}
   \includegraphics[width=0.49\textwidth]{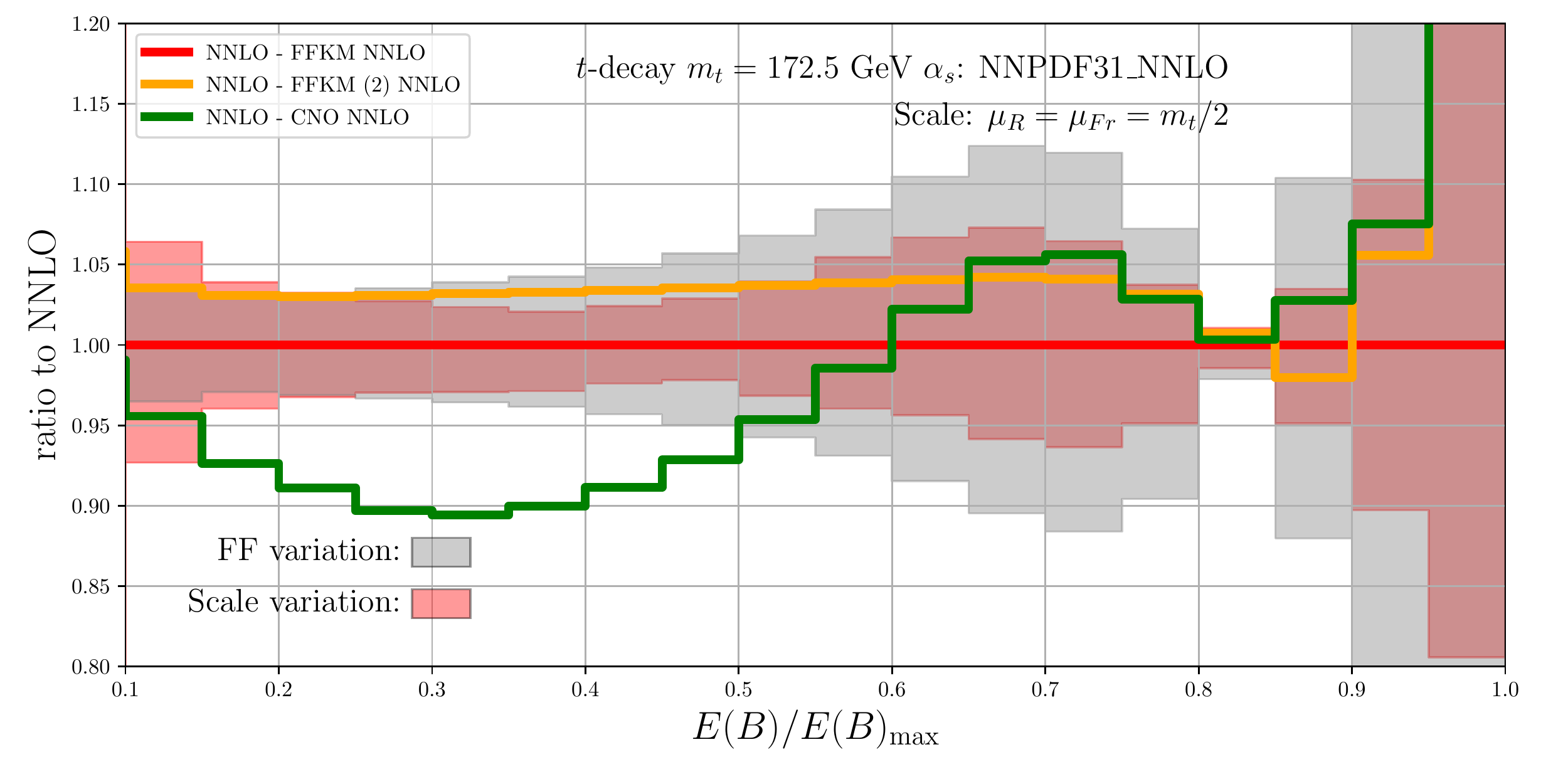}
  \caption{As in fig.~\ref{fig:top-decay-order} but showing the fragmentation function variation of the default FF FFKM at NNLO. Shown also are the central predictions for the other two FF at NNLO: FFKM(2) and CNO.}
  \label{fig:top-decay-fragmentation}
\end{figure}

The observables are shown in fig.~\ref{fig:top-decay-order}. In both cases we show the absolute distributions at different perturbative orders for the FFKM NNLO fragmentation function. The lower panel shows the ratio to the NLO result. The colored bands correspond to 7-point scale variation. In fig.~\ref{fig:top-decay-scale} we show a breakdown of the NNLO scale variation due to $\mu_R$ and $\mu_{Fr}$. Each one of these scales is varied (3-point variation) while the other scale is kept fixed at its central value. Similarly, fig.~\ref{fig:top-decay-fragmentation} shows the fragmentation function variation for the default FFKM fragmentation function at NNLO. Also shown are the central predictions at NNLO based on the other two FF sets: FFKM(2) and CNO.


The invariant mass differential width $m(B\ell)$ is of particular interest since it is suitable for extracting the top quark mass with high-precision \cite{Kharchilava:1999yj}. It has previously been studied with NLO precision in ref.~\cite{Biswas:2010sa}.

The normalized energy spectrum is also interesting in top mass determinations since it directly exposes the fragmentation function. Therefore it allows one to directly assess the sensitivity of this observable to $b$-fragmentation and its potential for measuring NPFF's. This observable has been studied in NLO+NLL QCD in \cite{Corcella:2001hz,Cacciari:2002re,Corcella:2005dk,Kniehl:2012mn,Nejad:2013fba,Nejad:2016epx}. The analytic expressions of the coefficient functions for both $m_b=0$ and $m_b\neq 0$ are known through NLO in QCD.

\subsection{$b$-fragmentation in top-quark pair-production and decay at the LHC}\label{sec:top-prod-and-decay}

In this section we present our predictions for the following $B$-hadron distributions in dilepton $t\bar t$ events at the LHC: the invariant mass of the $B$-hadron and charged lepton $m(B\ell)$ as well as $B$-hadron's energy $E(B)$. These two distributions are the $t\bar t$ equivalents of the distributions discussed in sec.~\ref{sec:top-decay} in the context of top quark decay. The advantage of working with $m(B\ell)$ and $E(B)$ is that they are defined in the detector frame and are, therefore, directly measurable without the need for reconstructing frames associated with the top quark. Both $m(B\ell)$ and $E(B)$ are of prime interest in the context of top quark mass determination at the LHC and have been extensively studied in the past in NLO QCD \cite{Kharchilava:1999yj,Biswas:2010sa,Agashe:2012bn,Agashe:2016bok}. 

The setup of the present calculation, which is closely related to the one in ref.~\cite{Czakon:2020qbd}, see also appendix \ref{sec:tt-xsec-fragmentation}, is as follows. We utilize the pdf set {\tt NNPDF3.1}. Its order is chosen in such a way that it matches the order of the perturbative calculation. The value of the strong coupling constant is obtained from the {\tt LHAPDF} library as provided by the {\tt NNPDF3.1} pdf set. The order of the strong coupling constant evolution in the perturbative calculation is matched to the order of the pdf while the order of the coupling in the FF evolution is matched to the order of the FF. 

The pdf variation utilizes the so-called reduced pdf set, see ref.~\cite{Czakon:2020qbd} for details. Our predictions are based on fixed central scales
\begin{equation}
\mu_R=\mu_F=\mu_{Fr}={m_t\over 2}\,.
\label{eq:scales}
\end{equation}

The reasons behind this scale choice are as follows. A fixed scale choice is well-justified in the kinematic ranges considered in this work. Furthermore, the use of fixed scales (instead of dynamic scales) can simplify the interpretation of the results especially when there are many scales and perturbative orders. The specific value of the central scale, $m_t/2$, is motivated by the study \cite{Czakon:2016dgf} on stable $t\bar t$ production. One may wonder if a central scale $m_t$ and not $m_t/2$ is more appropriate for the description of top decay. While both choices are equally suitable in principle and can be implemented in practice, we decided to use the scale choice (\ref{eq:scales}) in this first work on $b$-fragmentation in $t\bar t$ production and decay in order to make the interpretation of the scale variation of the prediction as transparent as possible since this way all three scales appearing in this calculation have the same central values. 

Scale variation is defined through a 15-scale variation, i.e. scaling up and down the {\it common} central scale by a factor of 2, subject to the constraints 
\begin{eqnarray}
&& 1/2 \le \mu_{R}/\mu_{F} \le 2\,, \nonumber\\
&& 1/2 \le \mu_{R}/\mu_{Fr} \le 2\,, \label{eq:scales-var}\\
&& 1/2 \le \mu_{F}/\mu_{Fr} \le 2\,.\nonumber
\end{eqnarray}
We use the $G_F$ scheme with the following parameters
\begin{eqnarray}
  m_W &=& 80.385 \; \text{GeV} \,,\nonumber\\ 
  \Gamma_W &=& 2.0928\;\text{GeV}\,,\nonumber\\
  m_Z &=& 91.1876 \; \text{GeV} \,,\nonumber\\ 
  \Gamma_Z &=& 2.4952\;\text{GeV}\,,\nonumber\\
  G_F &=& 1.166379\cdot10^{-5} \;\text{GeV}^{-2} \,,\nonumber\\ 
  \alpha &=&\frac{\sqrt{2} G_F}{\pi} m_W^2\left(1-(m_W/m_Z)^2\right)\,.
  \label{eq:GF-scheme}
\end{eqnarray}

Defining $\x=(m_W/m_t)^2$, the leading order top-quark width is computed from
\begin{equation}
\Gamma_t^{(0)} = G_F \frac{m_t^3}{8\pi\sqrt{2}}(1-\x)^2(1+2\x) = 1.48063 \;\text{GeV}~ ({\rm for}~ m_t = 172.5\;\text{GeV})\,.
\end{equation}

Our calculations are subject to typical phase space cuts: 
\begin{equation}
  p_T(B) \geq 10\, {\rm GeV} ~~~,~~~ |\eta(B)| \leq 2.4 \,.
  \label{eq:cuts}
\end{equation}
\begin{figure}[t]
  \centering
  \includegraphics[width=0.49\textwidth]{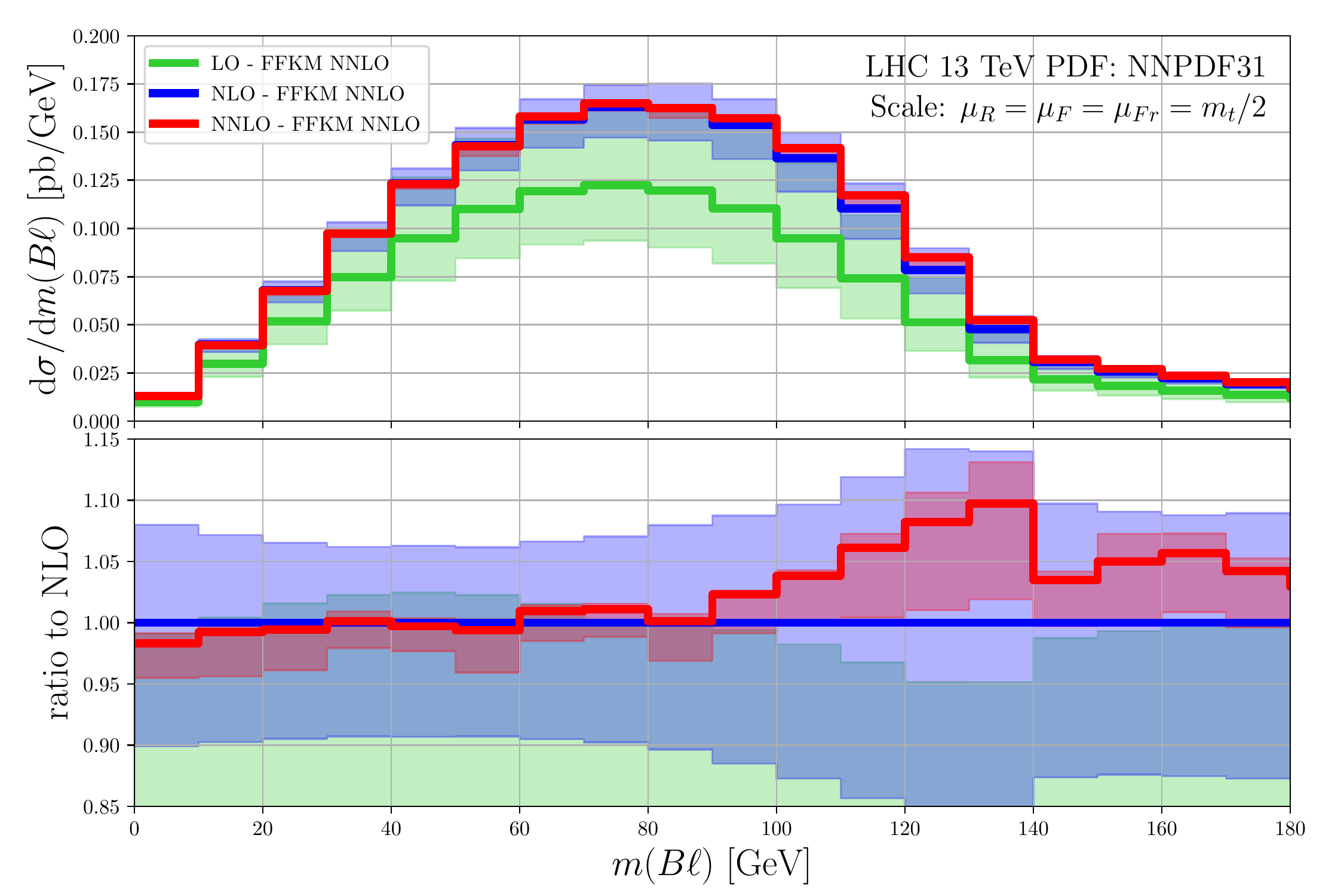}
  \includegraphics[width=0.49\textwidth]{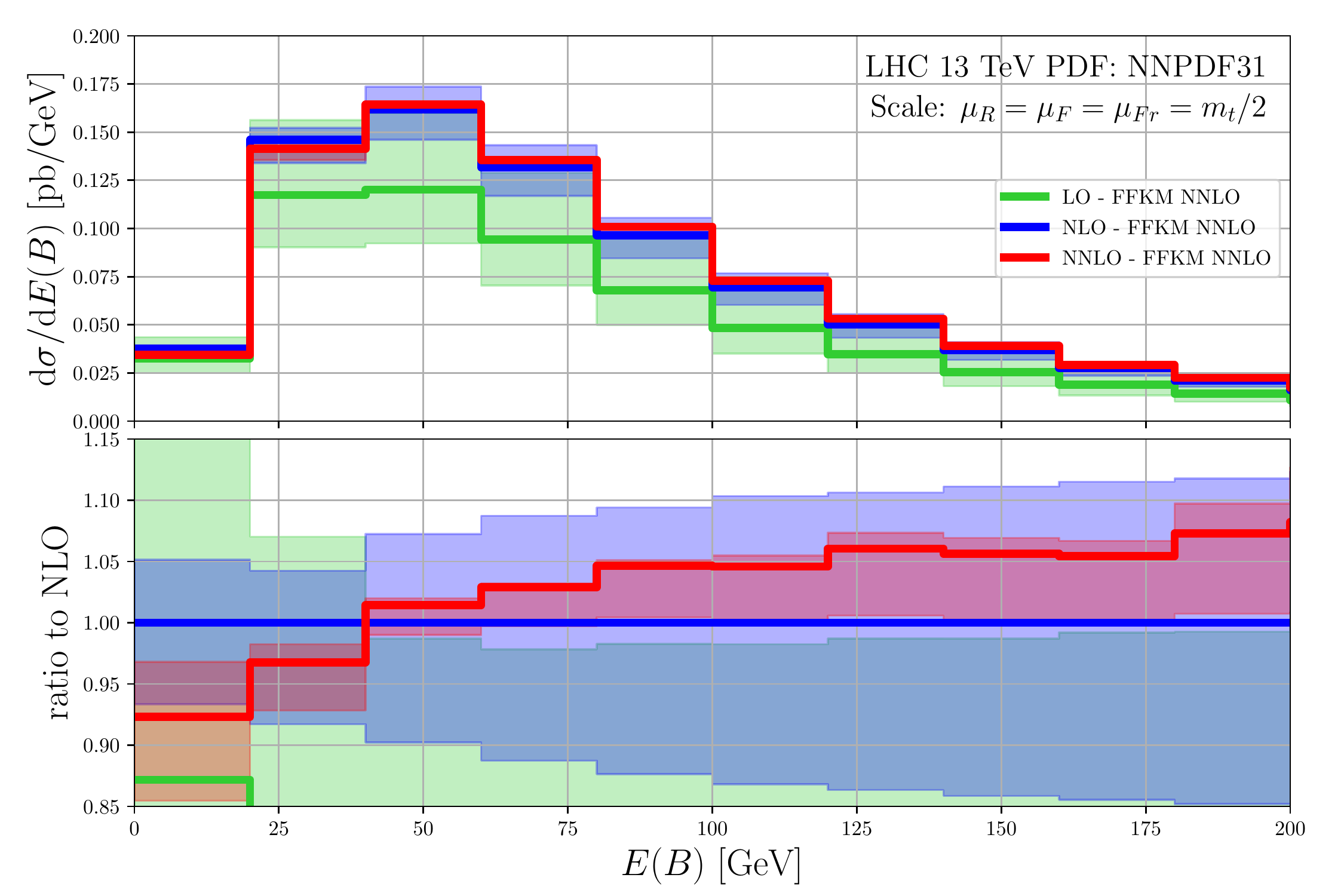}
  \caption{Absolute differential top-quark pair production and decay cross section as a function of the invariant mass $m(B\ell)$ (left) and the $B$-hadron energy $E(B)$ (right). All curves are convoluted with the same FF: FFKM at NNLO. Comparisons for LO, NLO and NNLO are shown.}
  \label{fig:ttbar-order}
\vskip 8mm
  \includegraphics[width=0.49\textwidth]{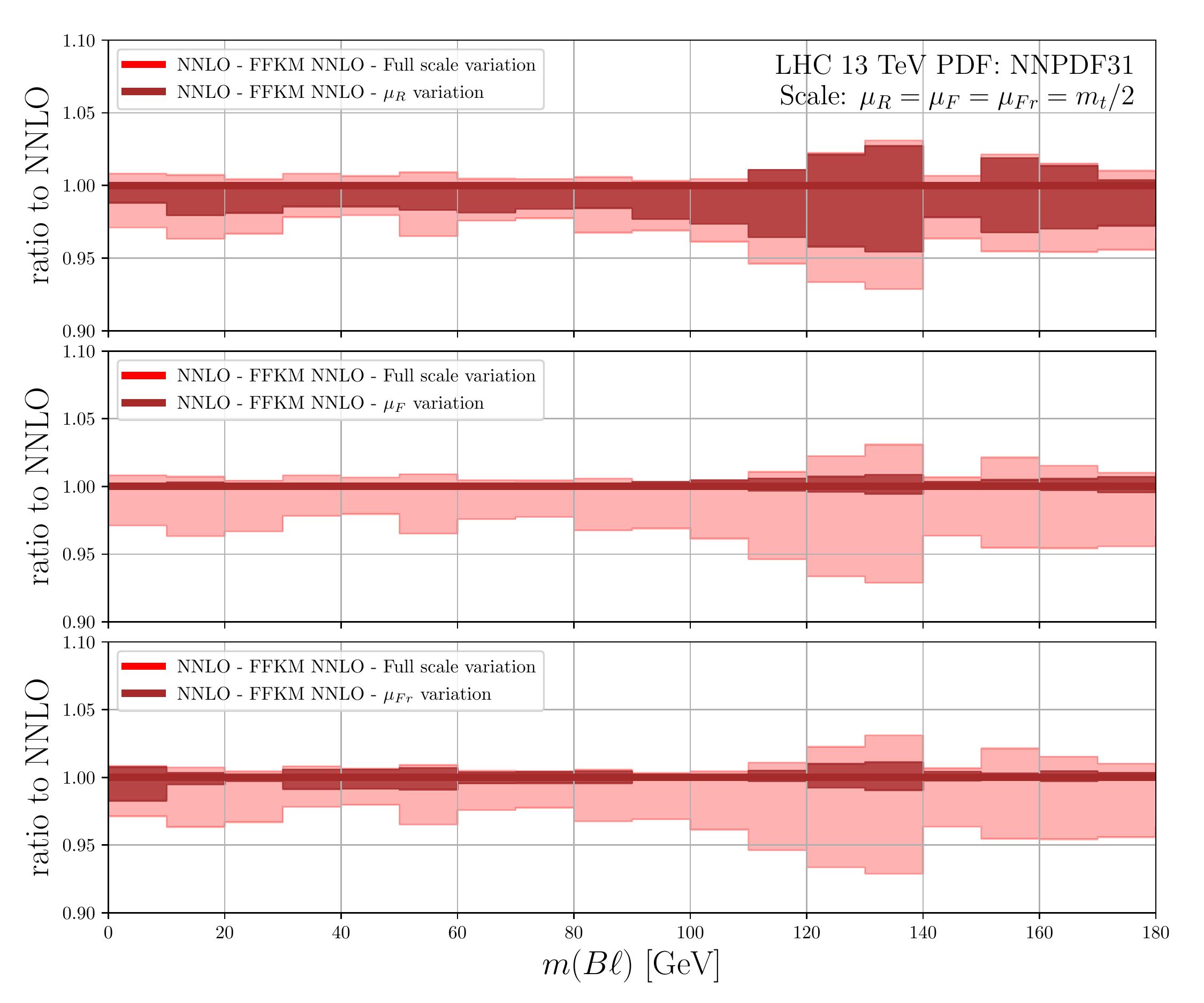}
  \includegraphics[width=0.49\textwidth]{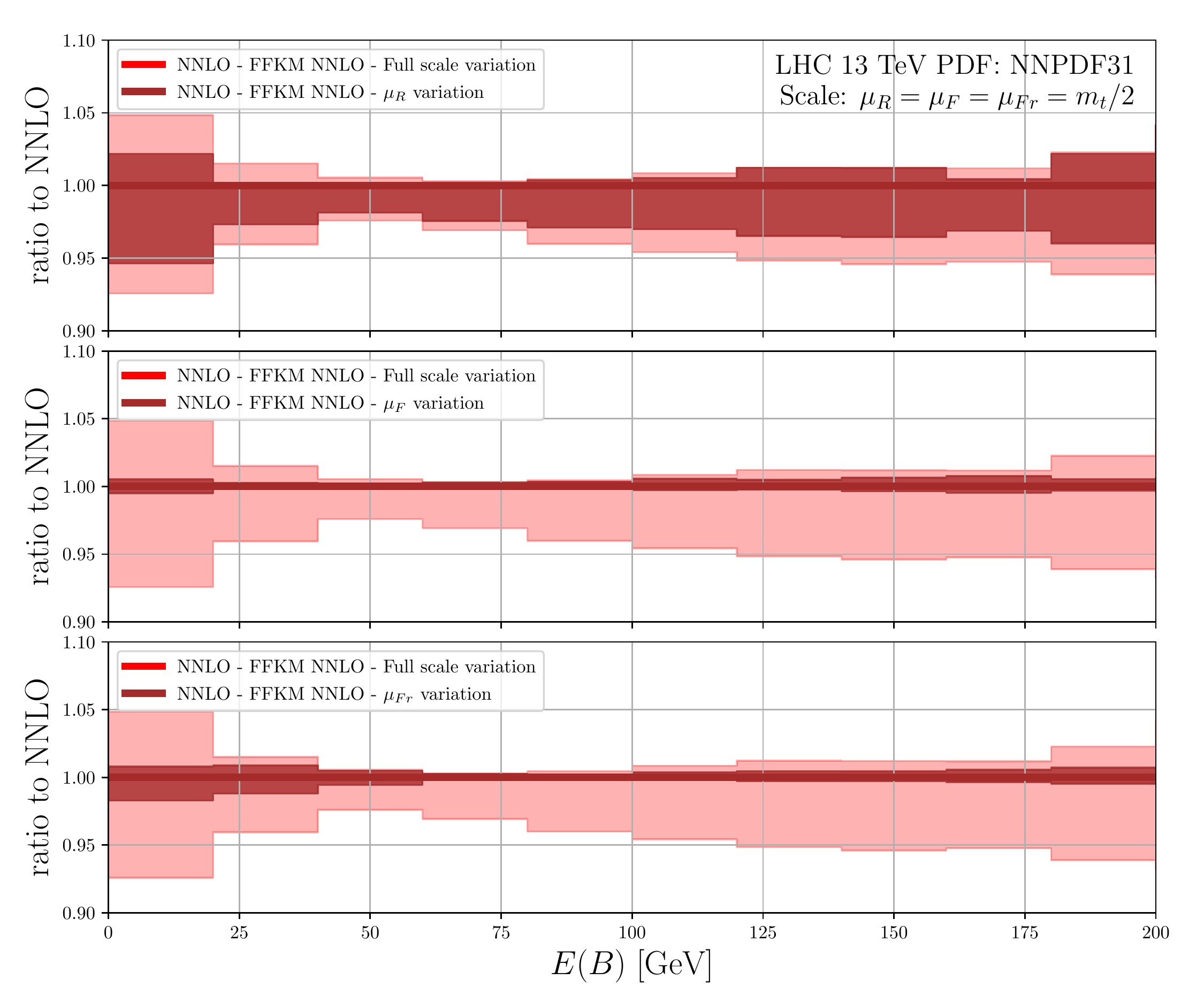}
  \caption{As in fig.~\ref{fig:ttbar-order} but showing the scale variation of the NNLO prediction:  $\mu_R$-only vs. total (upper plot), $\mu_F$-only vs. total (middle plot) and $\mu_{Fr}$-only vs. total scale variation (lower plot).}
  \label{fig:ttbar-scales}
\vskip 8mm
  \includegraphics[width=0.49\textwidth]{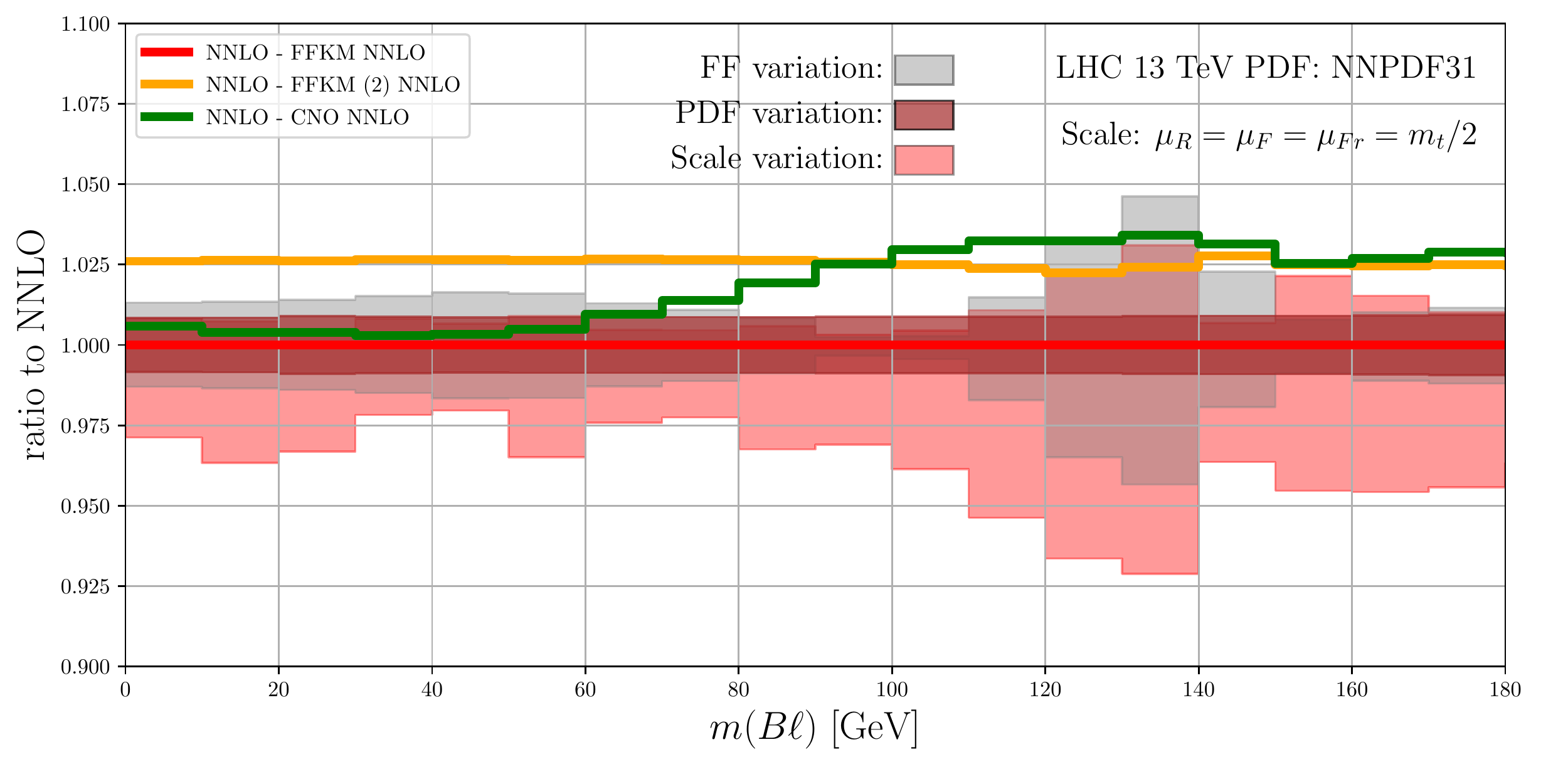}
  \includegraphics[width=0.49\textwidth]{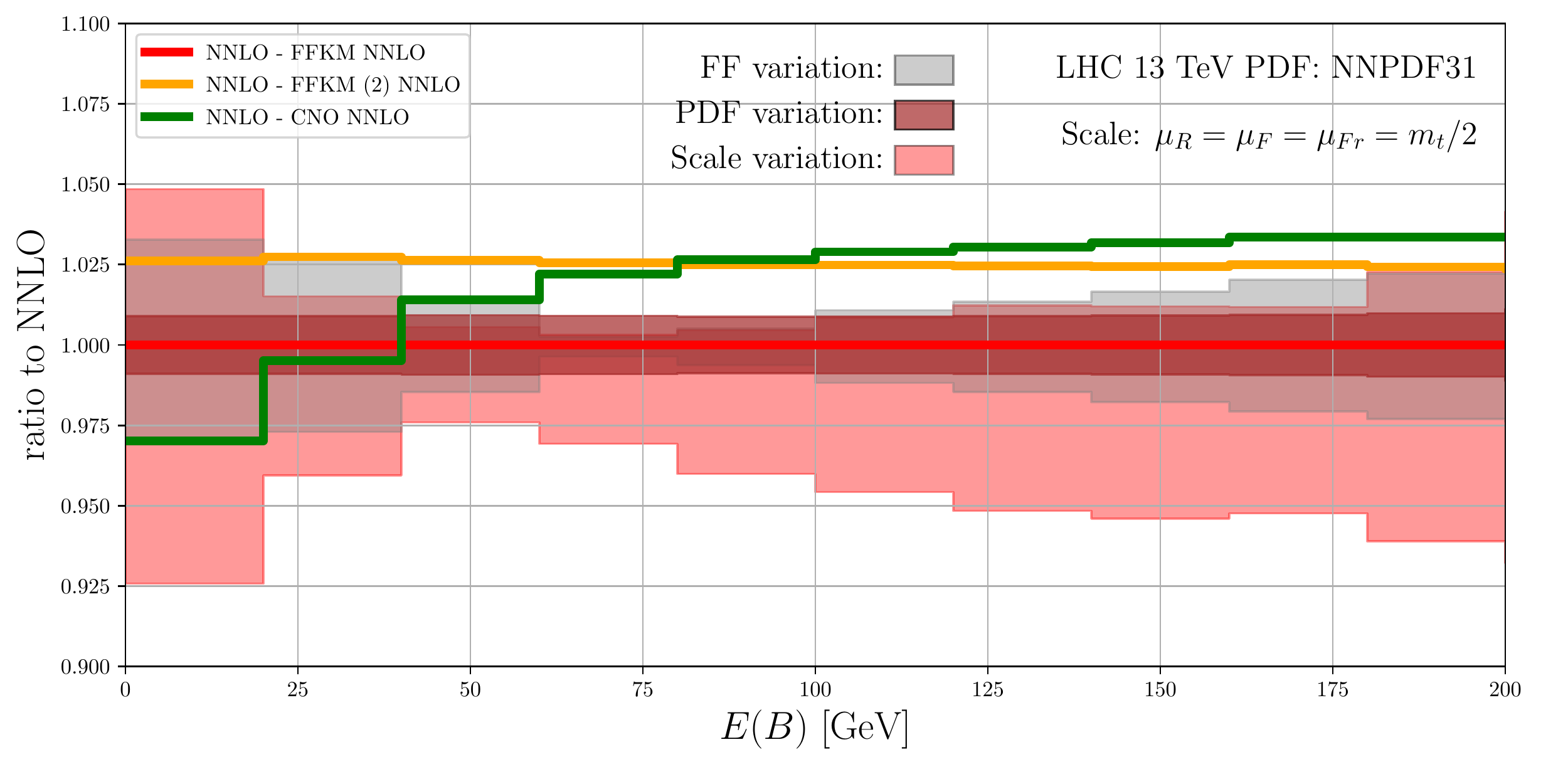}
  \caption{As in fig.~\ref{fig:ttbar-order} but showing the fragmentation and pdf variations of the default FFKM FF. Also shown are the central predictions for the other two FF at NNLO: FFKM(2) and CNO.}
  \label{fig:ttbar-fragmentation}
\end{figure}

The predictions for the absolute distributions $m(B\ell)$ and $E(B)$ through NNLO in QCD are shown in fig.~\ref{fig:ttbar-order}. The bands around the three central predictions indicate their 15-point scale variation. For both distributions we observe that the reduction of the scale uncertainty when going both from LO to NLO and from NLO to NNLO is substantial. The NNLO scale variation is about couple of percent in most bins. Notably, for the scale choice (\ref{eq:scales}) the NNLO scale variation is asymmetric, unlike the LO and NLO ones. Because of this asymmetry it is more useful to quantify the total width of the NNLO scale variation band which never exceeds 10\% and, in fact, in most bins is about half that value. This implies that the corrections due to missing higher order effects are probably at the one-percent level and thus rather small. 

We also observe that the size of the higher order corrections in both observables is moderate and in all cases the higher-order corrections are contained within the corresponding lower order scale variation band. The only exception is the lowest bin of the $E(B)$ distribution however it is worth keeping in mind that this bin is strongly impacted by the cuts (\ref{eq:cuts}). The NNLO/NLO K-factor is rather small and tends to be within $5$\% for most bins in both distributions. It has a non-trivial shape relative to the NLO predictions once one accounts for the small size of the NNLO scale uncertainty band. 

The region of the $m(B\ell)$ distribution above about 150 GeV is impacted by corrections beyond the narrow width approximation which is utilized in this work (see ref.~\cite{Czakon:2020qbd} for details). The monotonic increase in the shape of the NNLO/NLO $K$-factor of the $E(B)$ distribution suggests that at NNLO the maximum of that distribution is shifted towards higher values of $E(B)$ relative to NLO. Although in this paper we are not able to quantify this shift with sufficient  precision, we note that it may significantly affect any extraction of the top quark mass based on the proposal in refs.~\cite{Agashe:2012bn,Agashe:2016bok}. A more precise estimate of this effect is possible but it will require a dedicated and more refined calculation which we leave for a future work.

With the help of fig.~\ref{fig:ttbar-scales} one can assess the origin of the scale variation in these two observables at NNLO. To that end we have shown a breakdown of the scale variation due to one scale at a time (the other two being fixed at their central values) and compared to the total scale variation eq.~(\ref{eq:scales-var}). It immediately becomes apparent that the bulk of the scale variation is due to the renormalization scale $\mu_R$. The second largest contribution is due to the fragmentation scale $\mu_{Fr}$ while the contribution due to the factorization scale alone is tiny. 

In fig.~\ref{fig:ttbar-fragmentation} we compare at NNLO the three main sources of uncertainty for these two distributions: scale, pdf and fragmentation uncertainties. The variations shown are for the default FFKM fragmentation function. As an alternative measure for the fragmentation uncertainty we show the central predictions based on the two alternative FFs: FFKM(2) and CNO. It is evident from this figure that scale variation is the dominant source of uncertainty. This is true for all bins of both distributions. The second largest uncertainty is the one due to fragmentation followed by the pdf uncertainty. The differences between the three fragmentation functions tends to be consistent with the estimate of the fragmentation uncertainty although in some bins that difference is as large as twice the value of the fragmentation uncertainty estimate. 

In summary, the total uncertainty of the NNLO predictions for the $m(B\ell)$ and $E(B)$ distributions is within 5\% for almost all bins and is dominated by the scale uncertainty. While in this first NNLO work on this subject we have considered the 15-point scale variation eq.~(\ref{eq:scales-var}) around the central scale eq.~(\ref{eq:scales}) as the most straightforward generalization of the usual restricted scale variation in processes involving a single factorization scale, it may be beneficial to revisit this in the future and try to assess the impact and merits of a more restrictive scale variation and/or different dynamic or fixed scale choices.

\subsection{Extraction of $B$-hadron FFs from $t\bar t$ events}

The focus of the previous discussions was on predictions for LHC observables given a set of fragmentation functions. Due to the limitations of the existing extractions from $e^+e^-$ data one may naturally ask the question if LHC data can be used to improve the extraction of non-perturbative FFs. In this section we address this question in the context of $b$-fragmentation in $t\bar t$ events. As it will become clear shortly, this study can easily be extended to other processes like direct $b$ production. 

In principle, one can use any well-measured LHC $B$-hadron distribution to fit the NPFF. In order to increase the sensitivity to the NPFF it would be ideal if one uses distributions that are as closely related to the FF's as possible. An example for such a distribution is the $B$ energy spectrum in top quark decay discussed in sec.~\ref{sec:top-decay}. The only drawback of this distribution is that it requires the reconstruction of the decaying top quark and, thus, cannot be measured directly. It is therefore preferable to have distributions with similar sensitivity to NPFF that are directly defined in the lab frame. 

In this work we propose one such distribution: the ratio $p_T(B)/p_T(j_B)$ of the transverse momentum of the identified $B$-hadron with respect to the transverse momentum of the jet that contains it. We cluster jets with the anti-$k_T$ algorithm \cite{Cacciari:2008gp} with radius $R = 0.8$. We require that this jet fulfills $p_T(B) \leq p_T(j_B)$ and $|\eta(j_B)| < 2.4$, consistent with the cuts in eq.~(\ref{eq:cuts}). Note that both the $B$-hadron and its fragmentation remnants are included in this jet-clustering, see the discussion around eq.~(\ref{eq:splitting-kinematics}). 

\begin{figure}[t]
  \centering
  \includegraphics[width=0.5\textwidth]{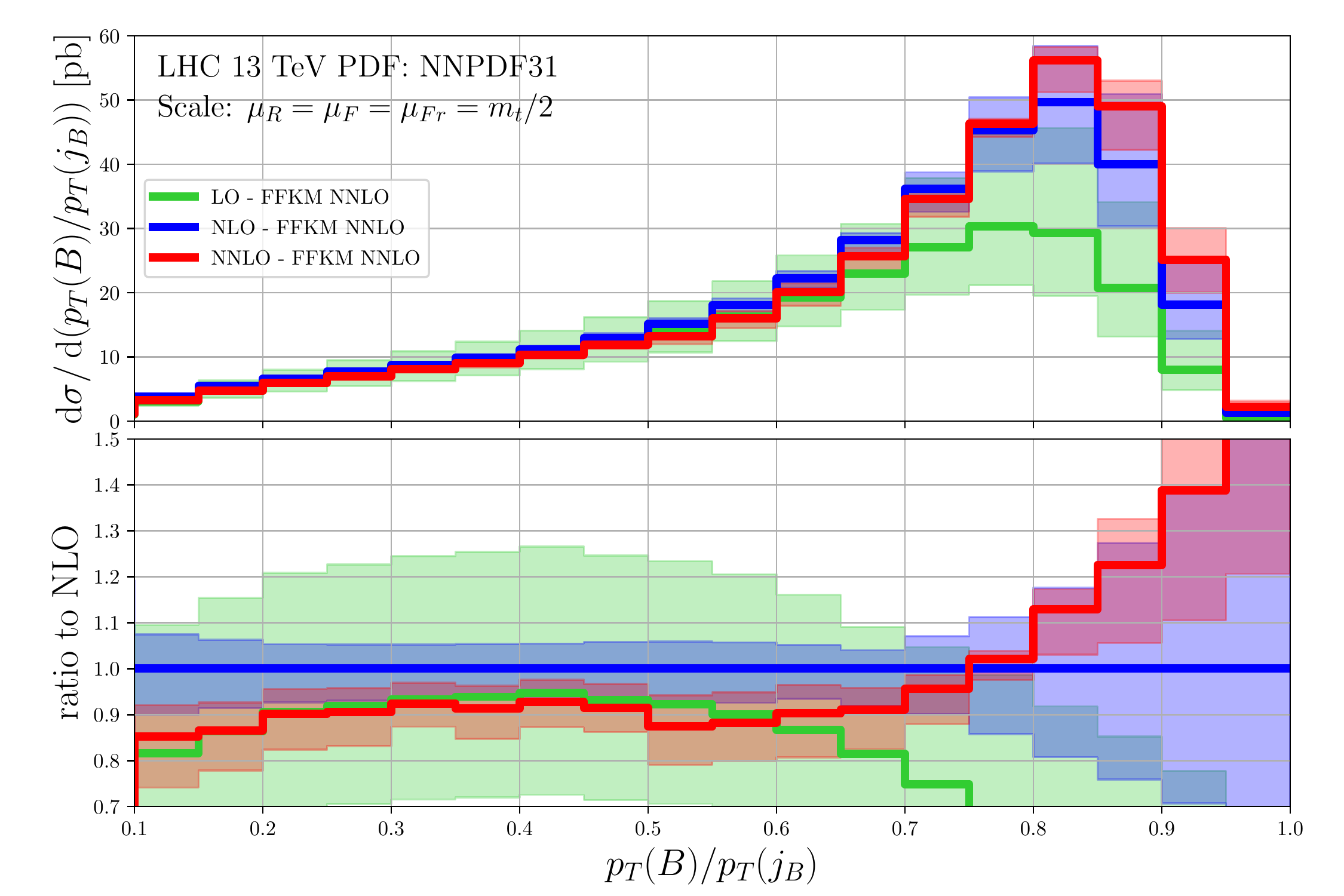}%
  \includegraphics[width=0.5\textwidth]{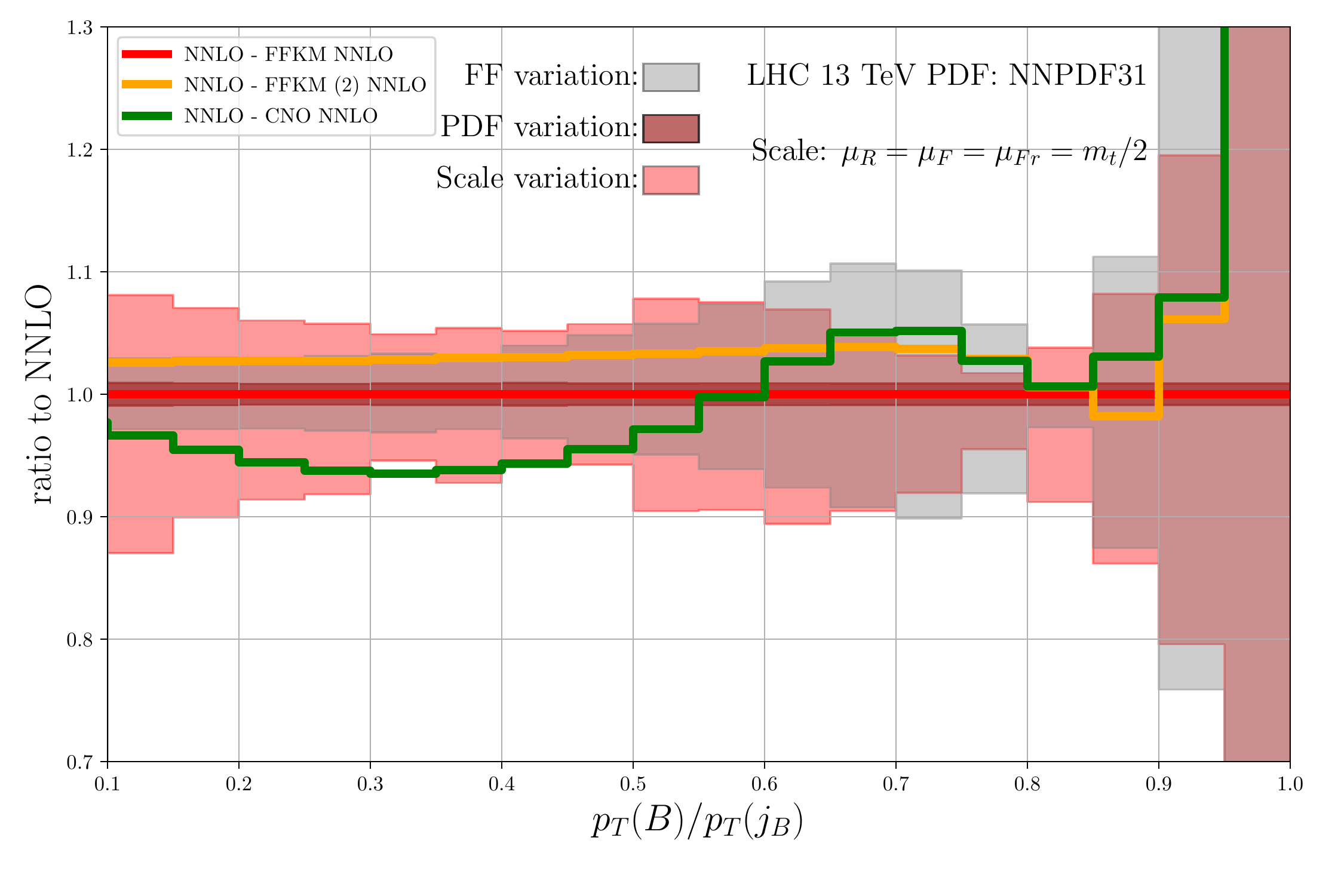}
  \caption{Absolute differential cross section as a function of the transverse momentum ratio $p_T(B)/p_T(j_B)$ in top-quark pair production and decay. Comparison of the FFKM NNLO FF for different perturbative orders showing scale variation (left) and comparison of FF, pdf and scale uncertainties (right). PDFs are matched to the corresponding perturbative order. The scale variation bands are based on 15-point scale variation.}
  \label{fig:prod_jetratio}
\end{figure}

The differential $p_T(B)/p_T(j_B)$ distribution is shown in fig.~\ref{fig:prod_jetratio}. The shape and behavior of this observable at different perturbative orders is fairly similar to the $E(B)/E(B)_{\rm max}$ distribution in top decay shown in fig.~\ref{fig:top-decay-order}. Higher order corrections are largely consistent with the scale uncertainty bands of the lower perturbative order. The size of scale variation at NNLO is below 5\% except for large values of $p_T(B)/p_T(j_B)$ where it starts to increase. We have checked that, just like in the case of $m(B\ell)$ and the $B$-hadron energy $E(B)$ distributions shown in fig.~\ref{fig:ttbar-scales}, the scale variation in this observable is driven by the renormalization scale and in much smaller degree, by the fragmentation scale $\mu_{Fr}$. The variation due to $\mu_F$ alone is negligible. 

From  fig.~\ref{fig:prod_jetratio} one can also conclude that for intermediate and large values of $p_T(B)/p_T(j_B)$ the uncertainty of this observable is driven by the uncertainty in the non-perturbative fragmentation function. For values $p_T(B)/p_T(j_B) \lesssim 0.5$ the total uncertainty is dominated by the scale variation. The pdf uncertainty is negligible throughout the kinematic range. These observations imply that this observable has strong potential for constraining FF at NNLO in QCD at intermediate and large values of $x$. 

We next probe the sensitivity of the $p_T(B)/p_T(j_B)$ distribution to the following parameters: the jet algorithm, the jet size and the $B$-hadron $\pt$ cut. Our aim is to determine optimal values for these parameters which will facilitate the extraction of the fragmentation function. 

\begin{figure}[t]
  \centering
    \includegraphics[width=0.49\textwidth]{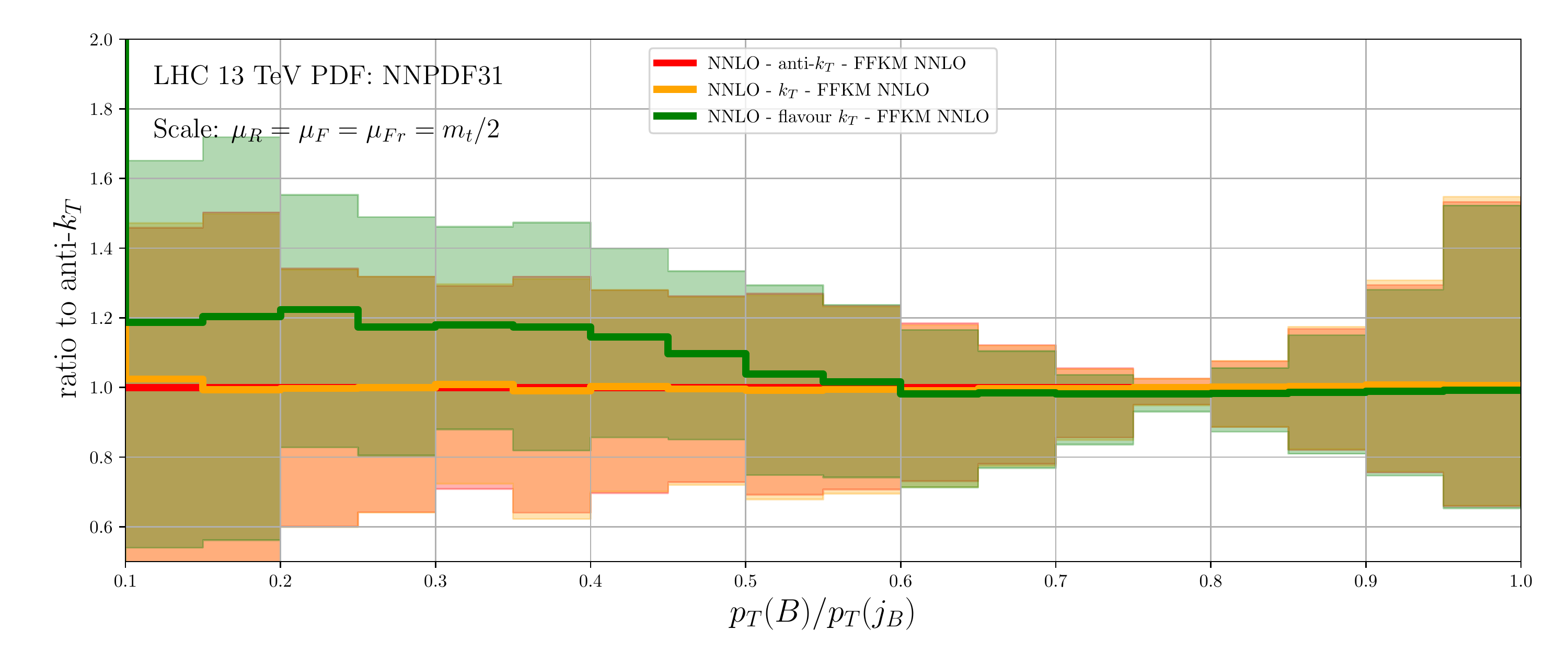}
    \includegraphics[width=0.49\textwidth]{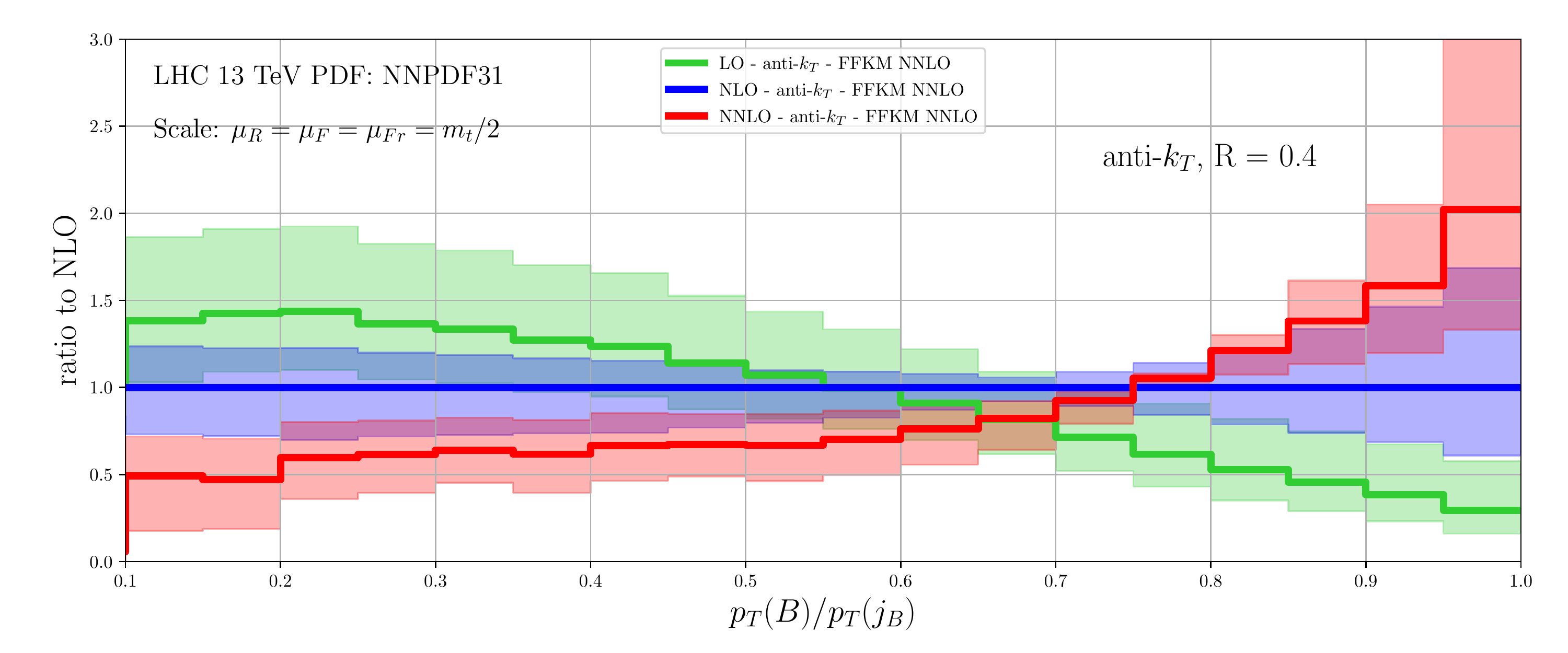}\\
    \includegraphics[width=0.49\textwidth]{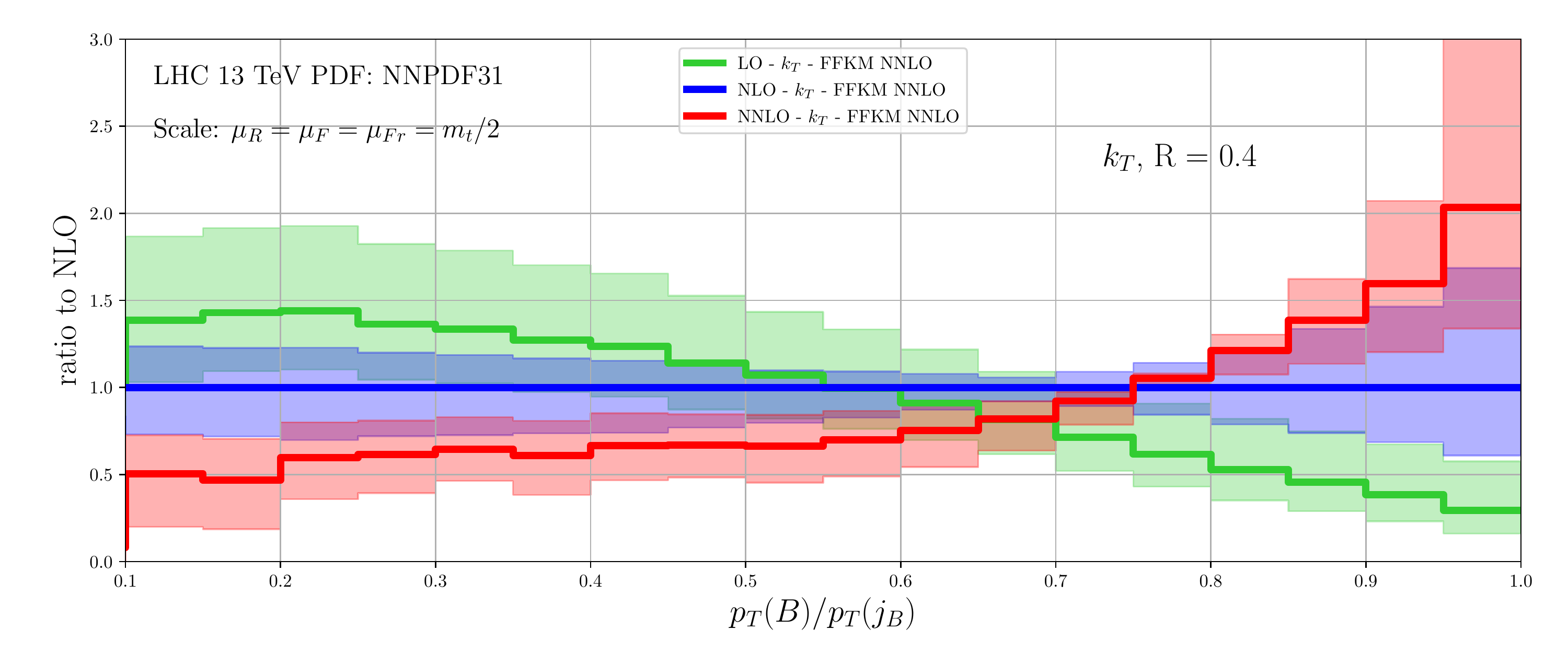}
    \includegraphics[width=0.49\textwidth]{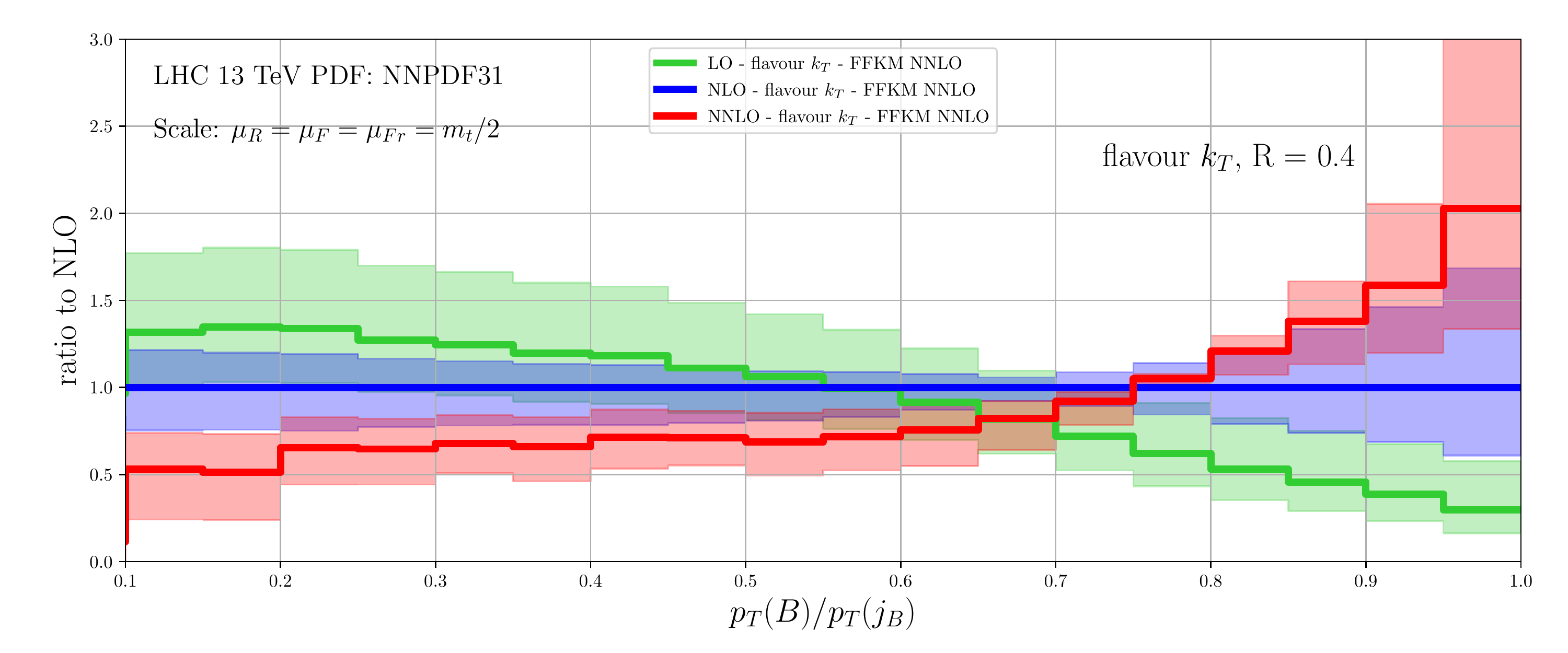}
  \caption{As in fig.~\ref{fig:prod_jetratio} but comparing different jet-algorithms: anti-$k_T$, $k_T$ and flavour-$k_T$.}
  \label{fig:prod_jetratio_jetalg}
\end{figure}

In fig.~\ref{fig:prod_jetratio_jetalg} we show the $p_T(B)/p_T(j_B)$ distribution for three different jet algorithms: anti-$k_T$, $k_T$ \cite{Catani:1993hr,Ellis:1993tq} and flavour-$k_T$ \cite{Banfi:2006hf}. For ease of the comparison all jet algorithms have the same jet size $R=0.4$. For each jet algorithm we show the LO, NLO and NNLO corrections, including their scale variation. The pattern of higher-order corrections is almost identical for the three jet algorithms. The three algorithms produce very similar distributions. This can be seen in the top left plot which shows a comparison of the three jet algorithms at NNLO. There we see that the anti-$k_T$ and $k_T$ algorithms lead to almost identical behavior. The flavour-$k_T$ algorithm also produces almost identical distribution for values of $p_T(B)/p_T(j_B)$ above about 0.6, but starts to deviate from the other two jet algorithms for lower values. Still the difference between the flavour-$k_T$ and the other two algorithms is much smaller than the NNLO scale uncertainty. These comparisons indicate that from the viewpoint of this observable all three jet algorithms, anti-$k_T$, $k_T$ and flavour-$k_T$, are suitable for the extraction of NPFF in $t\bar t$ events. 

Another comment about the use of the anti-$k_T$ and $k_T$ algorithms in this calculation is in order. It is well known \cite{Banfi:2006hf} that starting from NNLO, flavorless jet algorithms are not automatically infrared (IR) safe when applied to flavored problems. To achieve IR safety of jets in the flavored context, dedicated jet algorithms are needed. One such proposal is the flavour-$k_T$ algorithm of ref.~\cite{Banfi:2006hf}. Related ideas have been discussed in refs.~\cite{Buckley:2015gua,Dai:2018ywt}. 

The use of the anti-$k_T$ and $k_T$ algorithms is justified in the present work because of the special nature of the observables computed here. Unlike a typical fixed order calculation, in this work we cluster not just partons but the $B$-hadron and its accompanying remnants. Since by construction all collinear singularities have been regulated at the level of the partonic cross-section, a jet algorithm is no longer needed to ensure IR finiteness of the calculation. In this sense our calculation is closer to an experimental setup than to a typical fixed order partonic jet calculation. Since the fixed-order part of the $B$-hadron production cross-section contains terms of the type $\log^n(m)$ we expect that they will also be present in the corresponding jet calculation. However due to the NNLL DGLAP resummation they are likely to not play any role. 

\begin{figure}[t]
  \centering
  \includegraphics[width=0.49\textwidth]{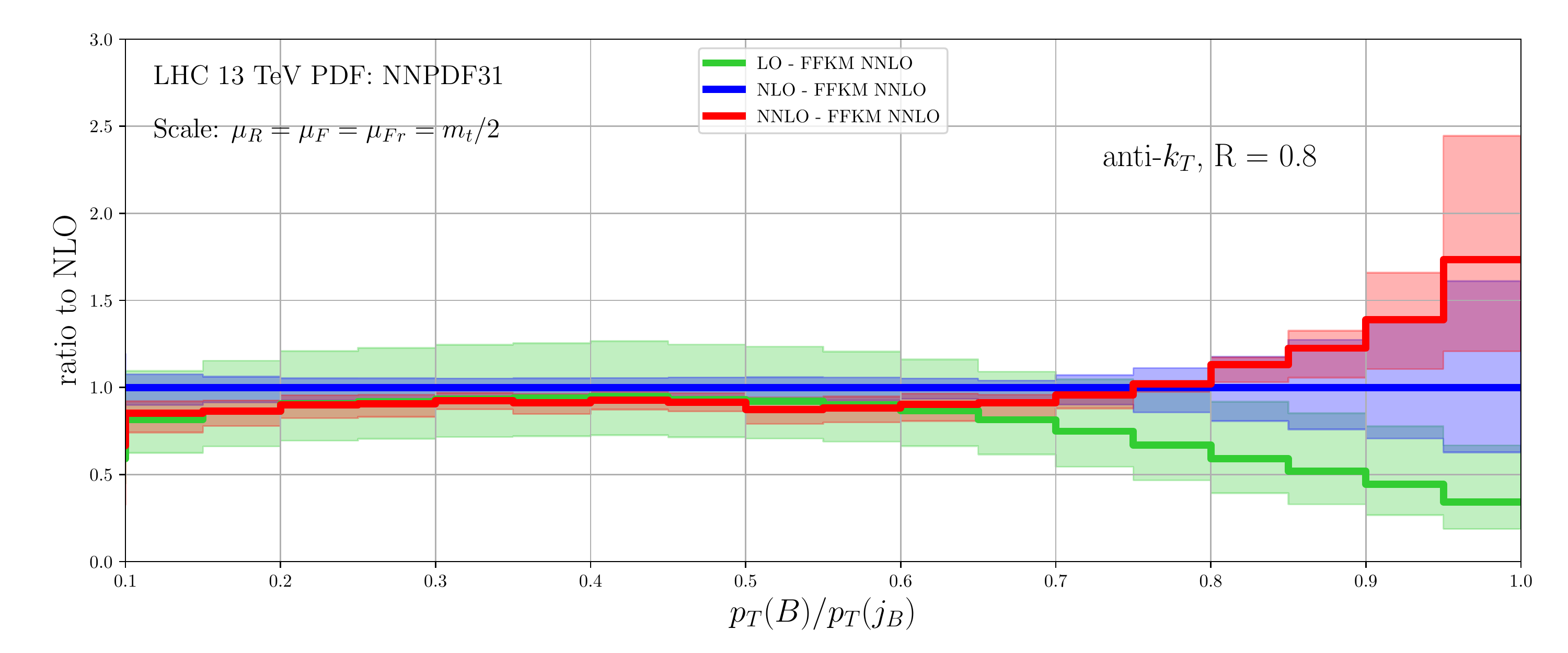}
  \includegraphics[width=0.49\textwidth]{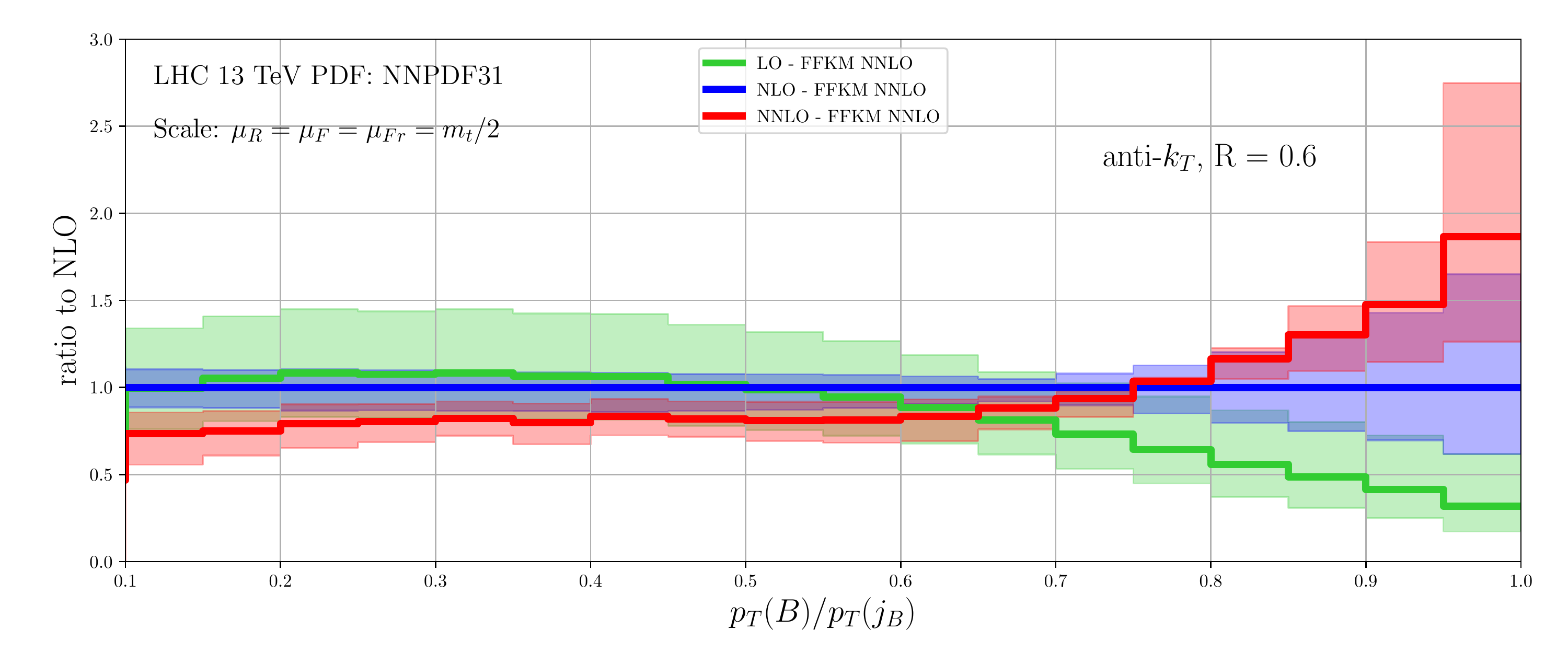}\\
  \includegraphics[width=0.49\textwidth]{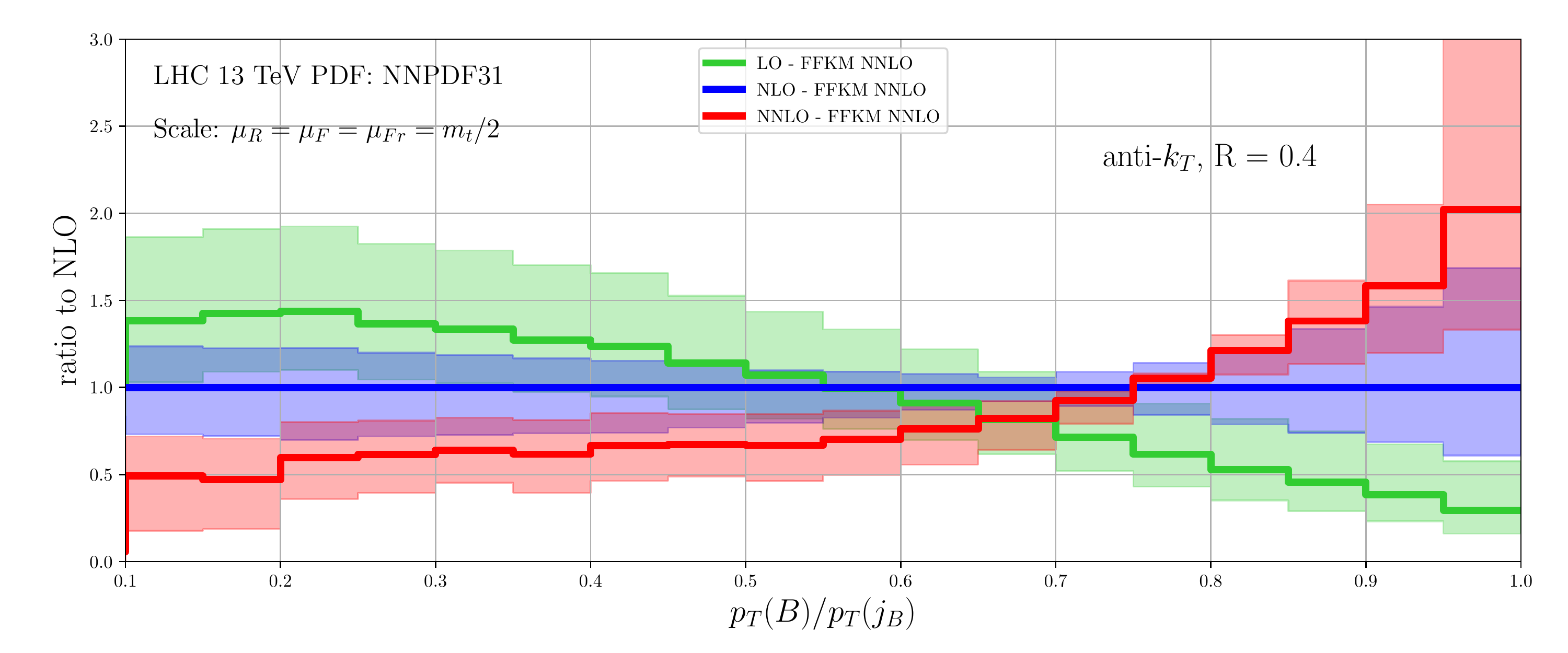}
  \includegraphics[width=0.49\textwidth]{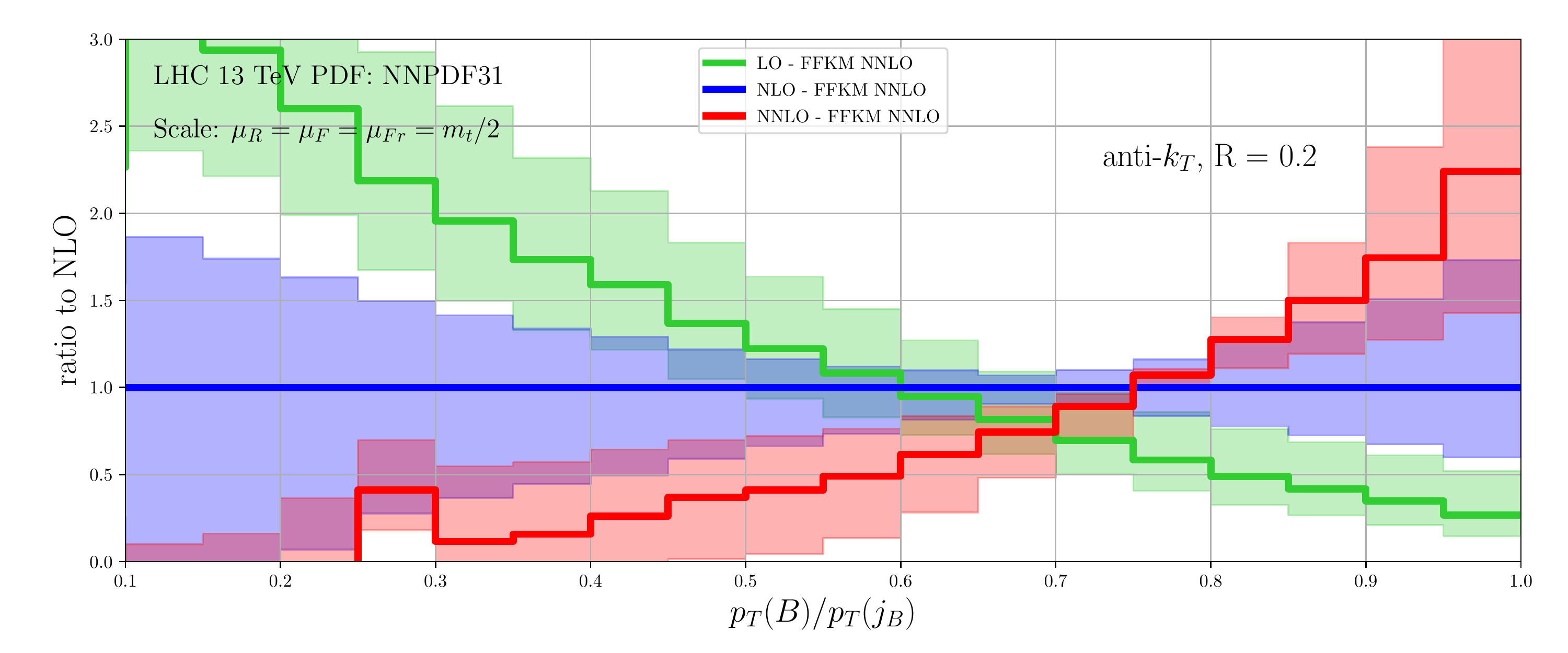} 
 \caption{As in fig.~\ref{fig:prod_jetratio} but comparing different jet sizes $R=0.2, 0.4, 0.6$ and $0.8$.}
  \label{fig:prod_jetratio_R}
  \end{figure}

We next consider the effect of the jet size $R$. In fig.~\ref{fig:prod_jetratio_R} we compare predictions based on the anti-$k_T$ algorithm with jet sized $R=0.2, 0.4, 0.6, 0.8$. We observe an expected pattern of higher order corrections: as the jet size decreases, the observable becomes less inclusive which results in decreased perturbative convergence. This is manifested through the increase of scale uncertainty at all orders considered in this calculation as well as larger $K$-factors. From this we concluded that from the viewpoint of theory, larger jet sizes are better for extracting fragmentation functions from the $p_T(B)/p_T(j_B)$ distribution.
\begin{figure}[t]
  \centering
  \includegraphics[width=0.49\textwidth]{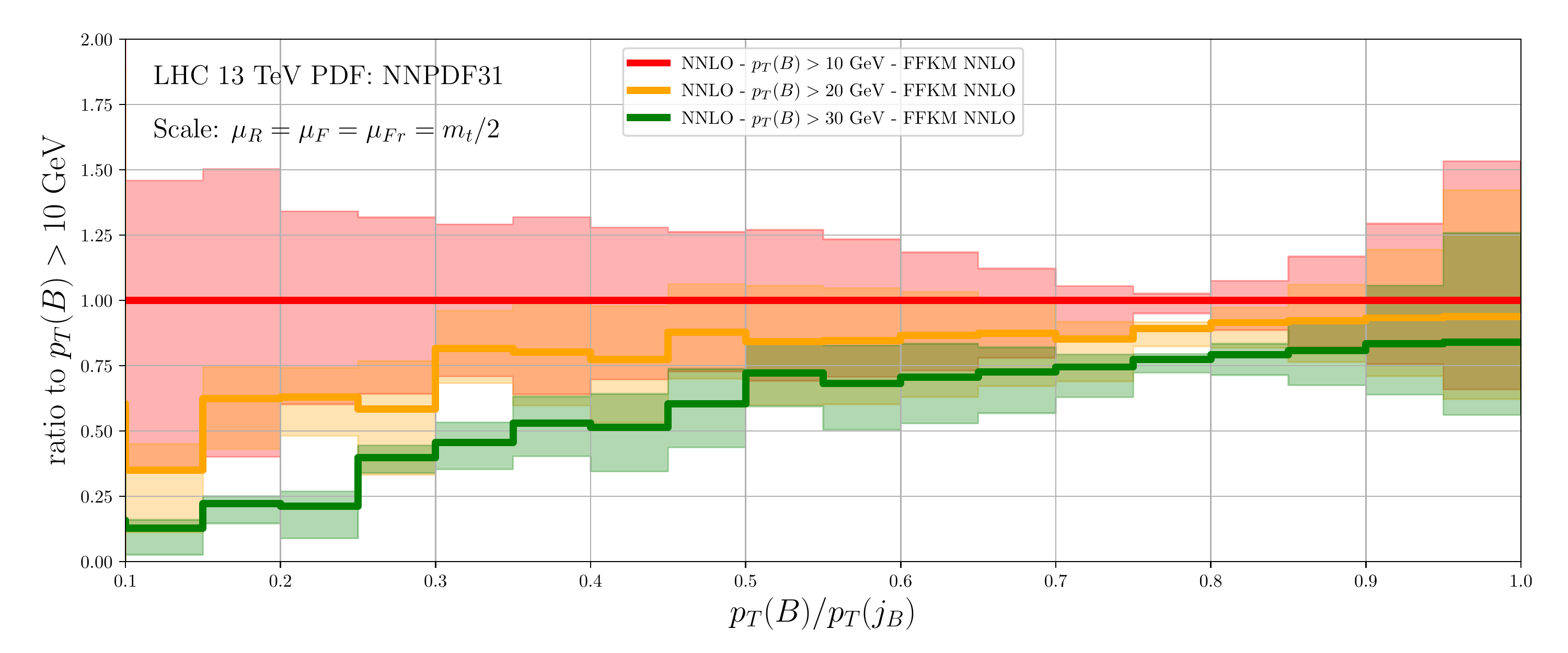}
  \includegraphics[width=0.49\textwidth]{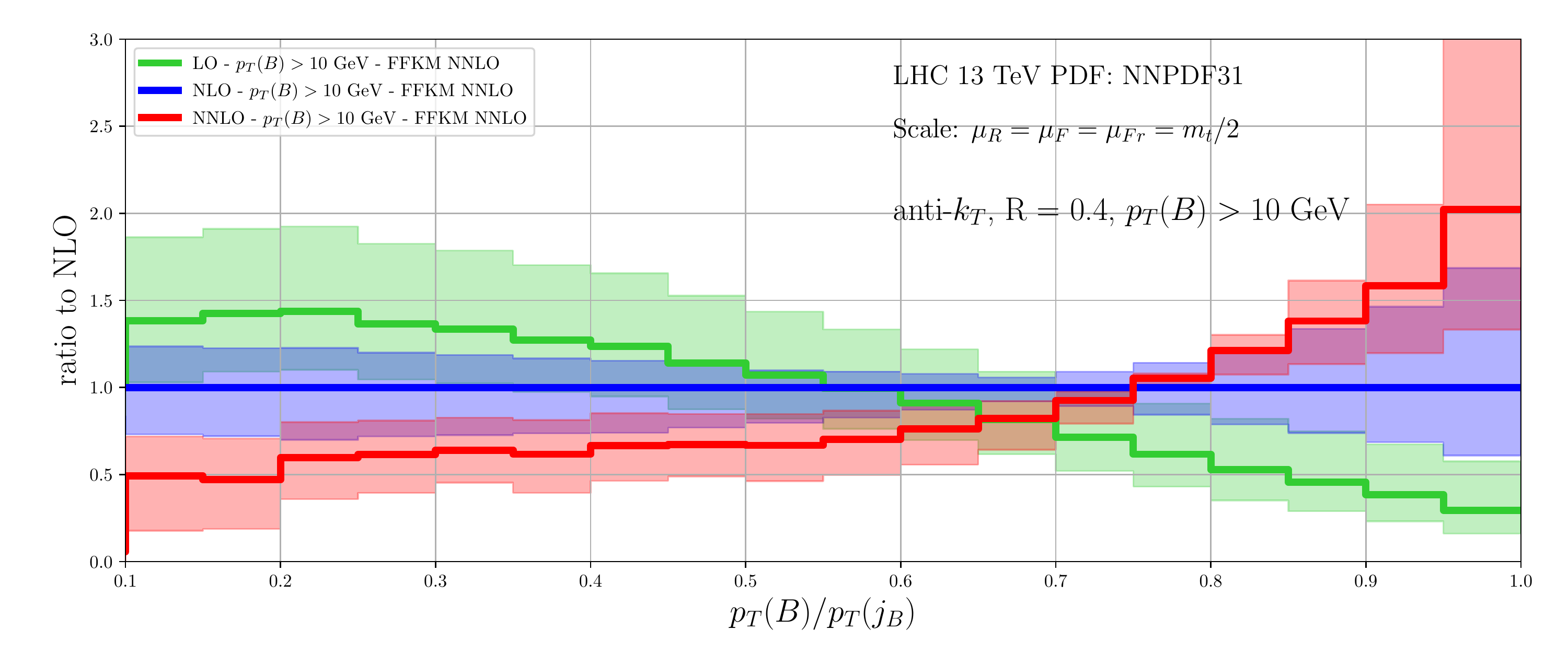}\\
  \includegraphics[width=0.49\textwidth]{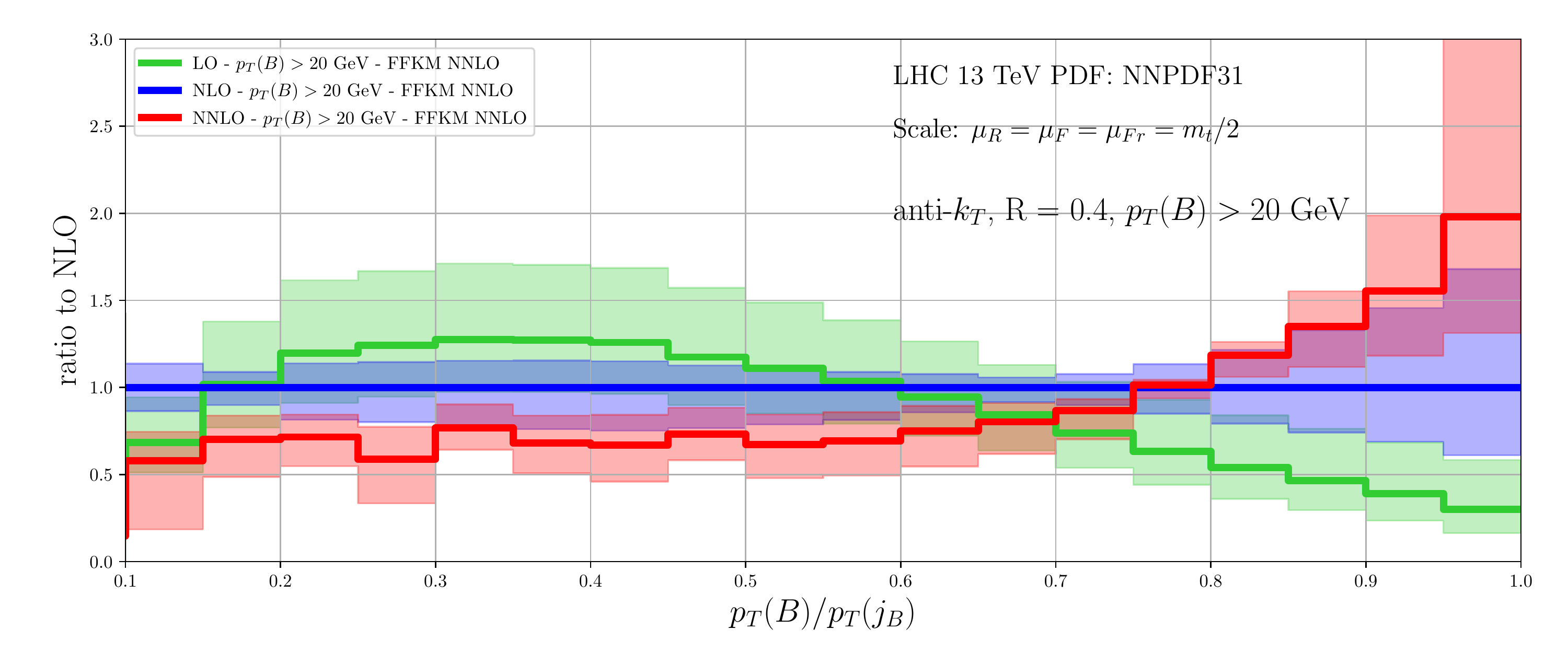}
  \includegraphics[width=0.49\textwidth]{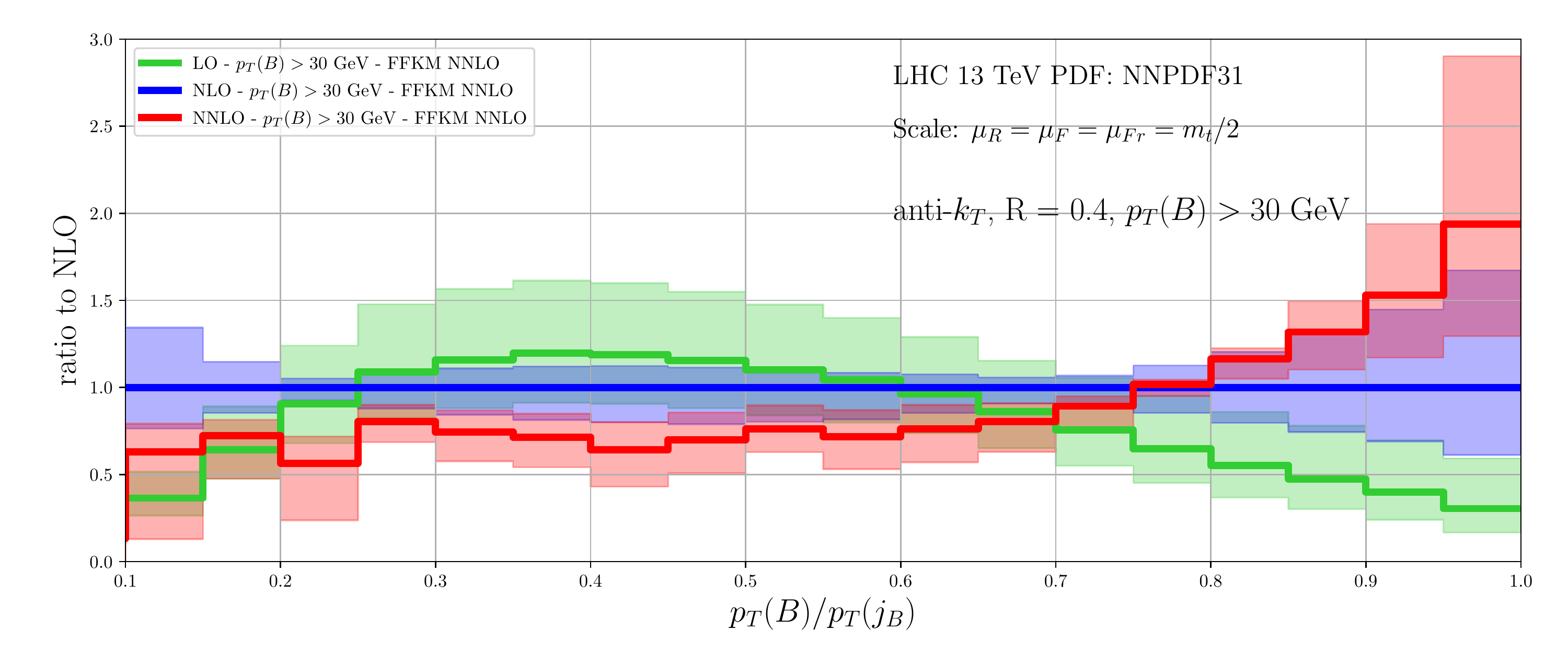}
 \caption{As in fig.~\ref{fig:prod_jetratio} but comparing different values of the $p_T(B)$ cut: $p_T(B) > 10, 20, 30$ GeV.}
  \label{fig:prod_jetratio_PTcut}
\end{figure}

Finally, we consider the impact of the low $p_T(B)$ cut. To that end in fig.~\ref{fig:prod_jetratio_PTcut} we show the $p_T(B)/p_T(j_B)$ distribution computed for three different values of this cut: $10, 20$ and $30$ GeV. We show the LO, NLO and NNLO distribution for each $p_T(B)$-cut as well as a comparison of the three cuts at NNLO. In all cases we use same jet algorithm: anti-$k_T$ with $R=0.4$. We observe that the intermediate-to-large $p_T(B)/p_T(j_B)$ region is not very much affected by the value of the low $p_T(B)$ cut which, in turn, means that the extracted fragmentation function at intermediate or large values of $x$ is not very sensitive to this cut. From the top-left plot in fig.~\ref{fig:prod_jetratio_PTcut} we observe that in this region the NNLO scale variation for all cut values is approximately the same. 

On the other hand, the value of the cut has a strong impact on the distribution at low $p_T(B)/p_T(j_B)$. As the $p_T(B)$ cut is lowered, the distribution becomes divergent in fixed order perturbation theory. This is consistent with the observed behavior of the distribution, which for smaller values of the $p_T(B)$ cut starts to show the typical signs of bad perturbative convergence: larger scale variation bands and increased $K$-factors. Finally, one should keep in mind that our calculation is performed with a massless $b$ quark and therefore misses corrections $\sim (m_b)^n$ for $n\geq 2$. For this reason it would be incomplete at low values of  $p_T(B)$. For these reasons we conclude that if experimentally viable, a larger $p_T(B)$ cut would be preferable since it leads to more stable predictions and since any missing $b$ mass corrections are automatically rendered negligible or at least significantly reduced in importance.

\section{Conclusions}\label{sec:conclusions}

Heavy flavor production at hadron colliders has traditionally demanded improved theoretical precision which matches the large statistics accumulated at colliders like the Tevatron and the LHC. In processes like $b$ and $c$ production, identified $b$- or $c$-flavored hadrons are copiously produced with transverse momenta much larger than their masses. For such kinematics the heavy quark mass plays the role of an infrared regulator. In an appropriately defined formalism, like the perturbative fragmentation function one we utilize in the present work, such mass effects could be consistently neglected.

In this work we extend for the first time the PFF formalism at hadron colliders to NNLO QCD. The novelty of the present work is that it develops a general, numeric, fully-flexible computational framework for perturbative cross sections for hadron collider processes with identified hadrons in NNLO QCD. Our work also benefits from the fact that all process-independent contributions needed for the description of heavy flavor fragmentation in NNLO -- like perturbative fragmentation functions, splitting functions and extracted from data non-perturbative fragmentation functions -- are available in the literature. Our framework is able to compute fully differential distributions with a single identified heavy hadron plus additional jets and non-strongly interacting particles. As a first application we compute the NNLO QCD corrections to $B$-hadron production in $t\bar t$ production with dilepton decays. The predicted realistic differential distributions significantly benefit from the inclusion of the NNLO QCD corrections. 

There are a number of ways the current work can be extended and we plan to pursue those in the near future. For example, one can compute open $B$ production at high $\pt$. The framework developed here can be extended in a straightforward way to charm production as well. 

One of the bottlenecks in this approach is the availability of high-quality non-perturbative fragmentation functions. These have previously been extracted from $e^+e^-$ data but the precision is not on par with current demand. In addition, the existing fragmentation functions are not fully compatible with our approach. To correct for this we intend to extract in the future non-perturbative fragmentation functions from $e^+e^-$ data within our framework.

In this work we have also studied the prospect of using LHC data for extracting $B$-hadron fragmentation functions. To that end we have proposed, and studied in detail, a distribution which we find to be particularly well suited for this task: the ratio of the $\pt$ of the $B$ hadron to the $\pt$ of the jet containing it. In the course of this study we have paid particular attention to the thorny problem of flavored jets in NNLO QCD.

Finally, an all-encompassing description of heavy flavor production in NNLO QCD will require the merging of fixed order calculations at low $\pt$ with the high $\pt$ description considered here. It is perhaps not too hard to envisage such a solution which, for example, builds on the FONLL approach at NLO. NNLO calculations with full mass dependence are possible as was recently demonstrated in ref.~\cite{Catani:2020kkl}. While such a merging is beyond the scope of the present work it represents a natural future extension of the present work.

\begin{acknowledgments}
The work of M.C. was supported by the Deutsche Forschungsgemeinschaft under grant 396021762 - TRR 257. The work of T.G. was supported by the Deutsche Forschungsgemeinschaft (DFG) under grant 400140256 - GRK 2497: The physics of the heaviest particles at the Large Hadron Collider. The research of A.M. and R.P. has received funding from the European Research Council (ERC) under the European Union's Horizon 2020 Research and Innovation Programme (grant agreement no. 683211). A.M. was also supported by the UK STFC grants ST/L002760/1 and ST/K004883/1.
\end{acknowledgments}

\appendix

\section{Structure of the cross section for $t\bar t$ production and top-quark decay including fragmentation}\label{sec:tt-xsec-fragmentation}

In the narrow-width approximation for the top quark, the differential cross-section for $t\bar t$ production and decay factorizes into three sub-processes: the top-pair production differential cross section and the differential widths for the top quark and antiquark
\begin{equation}
d\sigma = d\sigma_{t\overline{t}}\times\frac{d\Gamma_{t}}{\Gamma_t}\times\frac{d\Gamma_{\overline{t}}}{\Gamma_t}\,,
\label{eq:NWA-tt}
\end{equation}
where $\times$ denotes the properly accounted for spin correlations between the various factorized sub-processes. Through NNLO in QCD the three sub-processes can be expanded as follows
\begin{align}
d\sigma_{t\overline{t}} &= d\sigma_{t\overline{t}}^{(0)}+\alpha_s d\sigma_{t\overline{t}}^{(1)}+\alpha_s^2 d\sigma_{t\overline{t}}^{(2)}\,,\\
d\Gamma_{t(\overline{t})} &= d\Gamma_{t(\overline{t})}^{(0)}+\alpha_s d\Gamma_{t(\overline{t})}^{(1)}+\alpha_s^2 d\Gamma_{t(\overline{t})}^{(2)}\,.
\end{align}
Further details about the structure of the cross-section eq.~(\ref{eq:NWA-tt}) can be found in ref.~\cite{Czakon:2020qbd}. 

In the presence of fragmentation, i.e. for the process $pp \to t\bar t+X\to B+X$, the cross section in eq.~(\ref{eq:NWA-tt}) is further split into contributions depending on the origin of the fragmenting parton:
\begin{equation}
d\sigma = d\sigma_{t\overline{t}}\times\frac{d\Gamma_{t\to B}}{\Gamma_t}\times\frac{d\Gamma_{\overline{t}}}{\Gamma_t}+d\sigma_{t\overline{t}}\times\frac{d\Gamma_{t}}{\Gamma_t}\times\frac{d\Gamma_{\overline{t}\to B}}{\Gamma_t}+d\sigma_{t\overline{t}B}\times\frac{d\Gamma_{t}}{\Gamma_t}\times\frac{d\Gamma_{\overline{t}}}{\Gamma_t}\,,
\end{equation}
where the subscript $B$ is introduced to explicitly label the sub-process which initiates the fragmentation into the hadron $B$.

The fragmenting contributions have the following expansions through NNLO in QCD
\begin{align}
d\sigma_{t\overline{t}B} &= \alpha_s d\sigma_{t\overline{t}B}^{(1)}+\alpha_s^2 d\sigma_{t\overline{t}B}^{(2)}\,,\\
d\Gamma_{t(\overline{t})\to B} &= d\Gamma_{t(\overline{t})\to B}^{(0)}+\alpha_s d\Gamma_{t(\overline{t})\to B}^{(1)}+\alpha_s^2 d\Gamma_{t(\overline{t})\to B}^{(2)}\,,
\end{align}
where:
\begin{align}
d\sigma_{t\overline{t}B}^{(n)} &= \sum_i d\sigma_{t\overline{t}i}^{(n)}\otimes D_{i\to B} ~~~{\rm for}~~~ n=1,2\;,\\
d\Gamma_{t(\overline{t})\to B}^{(n)} &= \sum_i d\Gamma_{t(\overline{t})\to i}^{(n)}\otimes D_{i\to B} ~~~{\rm for}~~~ n=0,1,2\,.
\end{align}

The type of parton $i$ in the above equations that can fragment onto the observed hadron $B$ depends on the perturbative order. At LO, for example, no additional partons are present in $d\sigma_{t\overline{t}}$ while the only parton present in the top quark (antiquark) decay is $b$ ($\bar b$). At higher orders also the gluon and other quark flavors start to contribute.

\section{Collinear counterterms for processes involving fragmentation}\label{sec:collinear}

Here we present the explicit expressions for the collinear counterterms required for the calculation through NNLO QCD of any hadron collider process with fragmentation. The results below follow the conventions of ref.~\cite{Czakon:2014oma} and generalize the corresponding expressions given in that reference to processes involving fragmentation.

The NLO collinear renormalisation contribution reads
\begin{align}
\hat{\sigma}_{ab\to f_1...f_m[...]}^\text{C}&(p_1,p_2,k_1,...,k_m)=\notag\\
&\:\;\;\;\;\frac{\alpha_s}{2\pi}\frac{1}{\epsilon}\sum_c\int_0^1dz\Bigg[\left(\frac{\mu_R^2}{\mu_F^2}\right)^{\epsilon}P_{ca}^{(0)}(z)\hat{\sigma}_{cb\to f_1...f_m[...]}^\text{B}(zp_1,p_2,k_1,...,k_m)\notag\\
&\MathSpace+\left(\frac{\mu_R^2}{\mu_F^2}\right)^{\epsilon}P_{cb}^{(0)}(z)\hat{\sigma}_{ac\to f_1...f_m[...]}^\text{B}(p_1,zp_2,k_1,...,k_m)\notag\\
&\MathSpace+\frac{1}{z}\sum_i\left(\frac{\mu_R^2}{\mu_{Fr}^2}\right)^{\epsilon}P_{f_ic}^{(0)}(z)\hat{\sigma}_{ab\to f_1...c...f_m[...]}^\text{B}(p_1,p_2,k_1,...,k_i/z,...,k_m)\Bigg]\;,
\end{align}
where [...] represents non-fragmenting final state particles, $\mu_F$ is the PDF factorisation scale, $\mu_{Fr}$ is the fragmentation function factorisation scale and the relation $P_{ab}^{(0)\textrm{T}} = P_{ba}^{(0)\textrm{S}} \equiv P_{ba}^{(0)}$ has been used. The superscripts S and T in the splitting functions stand for space-like and time-like, respectively. 

The NNLO contributions read
\begin{align}
\hat{\sigma}_{ab\to f_1...f_m[...]}^\text{C1}&=
\frac{\alpha_s}{2\pi}\frac{1}{\epsilon}\sum_c\int_0^1dz\Bigg[\left(\frac{\mu_R^2}{\mu_F^2}\right)^{\epsilon}P_{ca}^{(0)}(z)\hat{\sigma}_{cb\to f_1...f_m[...]}^\text{R}(zp_1,...)\notag\\
&\MathSpace+\left(\frac{\mu_R^2}{\mu_F^2}\right)^{\epsilon}P_{cb}^{(0)}(z)\hat{\sigma}_{ac\to f_1...f_m[...]}^\text{R}(p_1,zp_2,...)\notag\\
&\MathSpace+\frac{1}{z}\sum_i\left(\frac{\mu_R^2}{\mu_{Fr}^2}\right)^{\epsilon}P_{f_ic}^{(0)}(z)\hat{\sigma}_{ab\to f_1...c...f_m[...]}^\text{R}(...,k_i/z,...)\Bigg]\;,
\end{align}
%
%
\begin{align}
\hat{\sigma}_{ab\to f_1...f_m[...]}^\text{C2}&=
\frac{\alpha_s}{2\pi}\frac{1}{\epsilon}\sum_c\int_0^1dz\Bigg[\left(\frac{\mu_R^2}{\mu_F^2}\right)^{\epsilon}P_{ca}^{(0)}(z)\hat{\sigma}_{cb\to f_1...f_m[...]}^\text{V}(zp_1,...)\notag\\
&\MathSpace+\left(\frac{\mu_R^2}{\mu_F^2}\right)^{\epsilon}P_{cb}^{(0)}(z)\hat{\sigma}_{ac\to f_1...f_m[...]}^\text{V}(p_1,zp_2,...)\notag\\
&\MathSpace+\frac{1}{z}\sum_i\left(\frac{\mu_R^2}{\mu_{Fr}^2}\right)^{\epsilon}P_{f_ic}^{(0)}(z)\hat{\sigma}_{ab\to f_1...c...f_m[...]}^\text{V}(...,k_i/z,...)\Bigg]\notag\\
&+\left(\frac{\alpha_s}{2\pi}\right)^2\frac{1}{2\epsilon}\sum_c\int_0^1dz\Bigg[\left(\frac{\mu_R^2}{\mu_F^2}\right)^{2\epsilon}P_{ca}^{(1)\textrm{S}}(z)\hat{\sigma}_{cb\to f_1...f_m[...]}^\text{B}(zp_1,...)\notag\\
&\MathSpace+\left(\frac{\mu_R^2}{\mu_F^2}\right)^{2\epsilon}P_{cb}^{(1)\textrm{S}}(z)\hat{\sigma}_{ac\to f_1...f_m[...]}^\text{B}(p_1,zp_2,...)\notag\\
&\MathSpace+\frac{1}{z}\sum_i\left(\frac{\mu_R^2}{\mu_{Fr}^2}\right)^{2\epsilon}P_{cf_i}^{(1)\textrm{T}}(z)\hat{\sigma}_{ab\to f_1...c...f_m[...]}^\text{B}(...,k_i/z,...)\Bigg]\notag\\
&+\left(\frac{\alpha_s}{2\pi}\right)^2\frac{\beta_0}{4\epsilon^2}\sum_c\int_0^1dz\Bigg[\left\{\left(\frac{\mu_R^2}{\mu_F^2}\right)^{2\epsilon}-2\left(\frac{\mu_R^2}{\mu_F^2}\right)^{\epsilon}\right\}P_{ca}^{(0)}(z)\hat{\sigma}_{cb\to f_1...f_m[...]}^\text{B}(zp_1,...)\notag\\
&\MathSpace+\left\{\left(\frac{\mu_R^2}{\mu_F^2}\right)^{2\epsilon}-2\left(\frac{\mu_R^2}{\mu_F^2}\right)^{\epsilon}\right\}P_{cb}^{(0)}(z)\hat{\sigma}_{ac\to f_1...f_m[...]}^\text{B}(p_1,zp_2,...)\notag\\
&\MathSpace+\frac{1}{z}\sum_i\left\{\left(\frac{\mu_R^2}{\mu_{Fr}^2}\right)^{2\epsilon}-2\left(\frac{\mu_R^2}{\mu_{Fr}^2}\right)^{\epsilon}\right\}P_{f_ic}^{(0)}(z)\hat{\sigma}_{ab\to f_1...c...f_m[...]}^\text{B}(...,k_i/z,...)\Bigg]\notag\\
&+\left(\frac{\alpha_s}{2\pi}\right)^2\frac{1}{2\epsilon^2}\sum_{cd}\int_0^1dz\Bigg[\left(\frac{\mu_R^2}{\mu_F^2}\right)^{2\epsilon}\left(P_{cd}^{(0)}\otimes P_{da}^{(0)}\right)(z)\hat{\sigma}_{cb\to f_1...f_m[...]}^\text{B}(zp_1,...)\notag\\
&\MathSpace+\left(\frac{\mu_R^2}{\mu_F^2}\right)^{2\epsilon}\left(P_{cd}^{(0)}\otimes P_{db}^{(0)}\right)(z)\hat{\sigma}_{ac\to f_1...f_m[...]}^\text{B}(p_1,zp_2,...)\notag\\
&\MathSpace+\frac{1}{z}\sum_i\left(\frac{\mu_R^2}{\mu_{Fr}^2}\right)^{2\epsilon}\left(P_{f_id}^{(0)}\otimes P_{dc}^{(0)}\right)(z)\hat{\sigma}_{ab\to f_1...c...f_m[...]}^\text{B}(...,k_i/z,...)\Bigg]\notag\\
&+\left(\frac{\alpha_s}{2\pi}\right)^2\frac{1}{\epsilon^2}\sum_{cd}\iint_0^1dzd\overline{z}\Bigg[\left(\frac{\mu_R^2}{\mu_F^2}\right)^{2\epsilon}P_{ca}^{(0)}(z)P_{db}^{(0)}(\overline{z})\hat{\sigma}_{cd\to f_1...f_m[...]}^\text{B}(zp_1,\overline{z}p_2,...)\notag\\
&\MathSpace+\frac{1}{\overline{z}}\sum_{i}\left(\frac{\mu_R^2}{\mu_{F}\mu_{Fr}}\right)^{2\epsilon}P_{ca}^{(0)}(z)P_{f_id}^{(0)}(\overline{z})\hat{\sigma}_{cb\to f_1...d...f_m[...]}^\text{B}(zp_1,...,k_i/\overline{z}, ...)\notag\\
&\MathSpace+\frac{1}{\overline{z}}\sum_{i}\left(\frac{\mu_R^2}{\mu_{F}\mu_{Fr}}\right)^{2\epsilon}P_{cb}^{(0)}(z)P_{f_id}^{(0)}(\overline{z})\hat{\sigma}_{ac\to f_1...d...f_m[...]}^\text{B}(p_1,zp_2,...,k_i/\overline{z}, ...)\notag\\
&\MathSpace+\frac{1}{z\overline{z}}\sum_{i< j}\left(\frac{\mu_R^2}{\mu_{Fr}^2}\right)^{2\epsilon}P_{f_ic}^{(0)}(z)P_{f_jd}^{(0)}(\overline{z})\hat{\sigma}_{ab\to f_1...c...d...f_m[...]}^\text{B}(...,k_i/z,...,k_j/\overline{z}, ...)\Bigg]\;,
\end{align}
where, for compactness, arguments without factors of $z$ or $\overline{z}$ have been omitted.

\section{Checks on our computational setup}

In the following we detail two checks of our NNLO calculational setup defined in sec.~\ref{sec:computational-framework}.

\subsection{$b$-fragmentation in $e^+e^-$ collisions}\label{sec:e+e-}

\begin{figure}[t]
  \centering
  \includegraphics[width=0.49\textwidth]{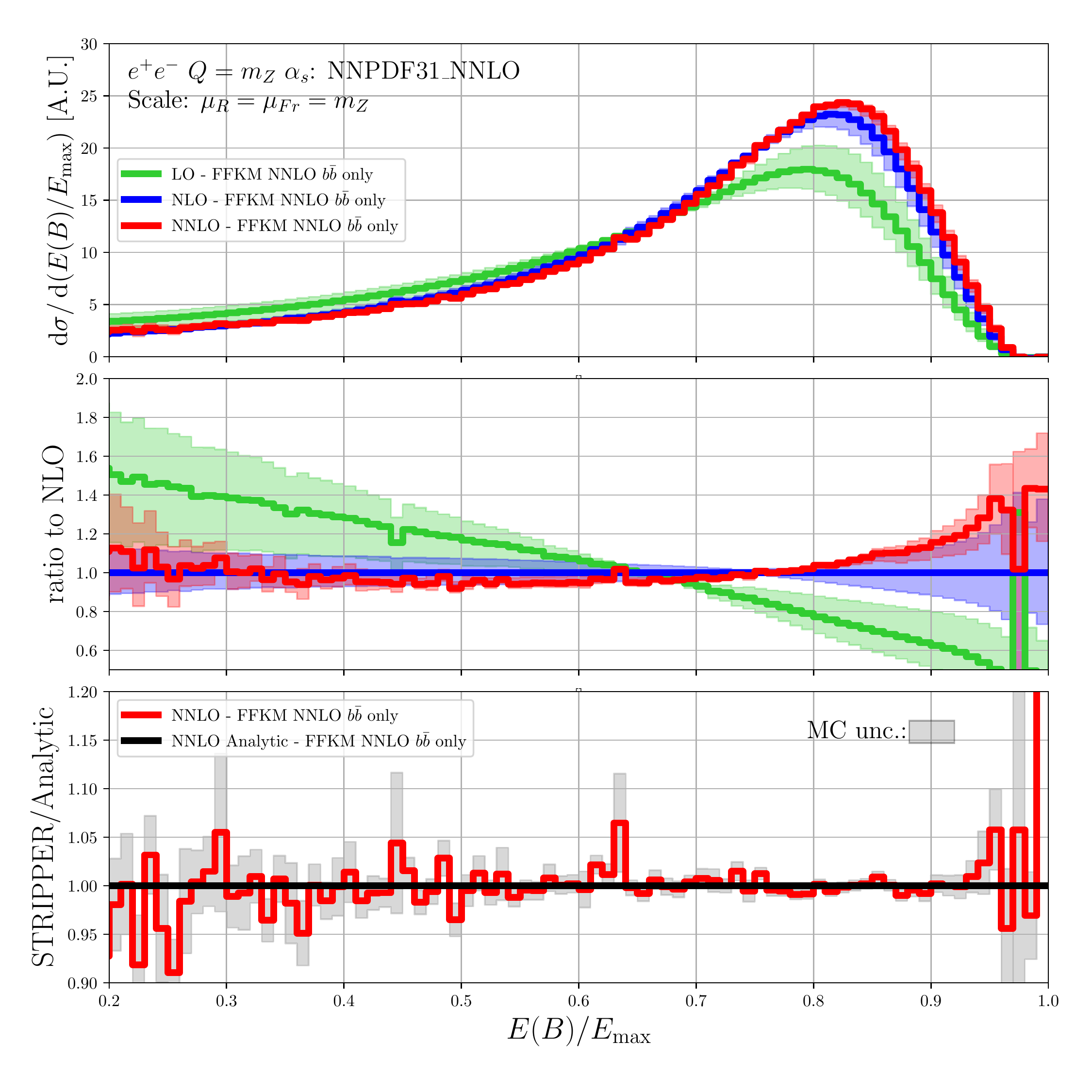}
  \includegraphics[width=0.49\textwidth]{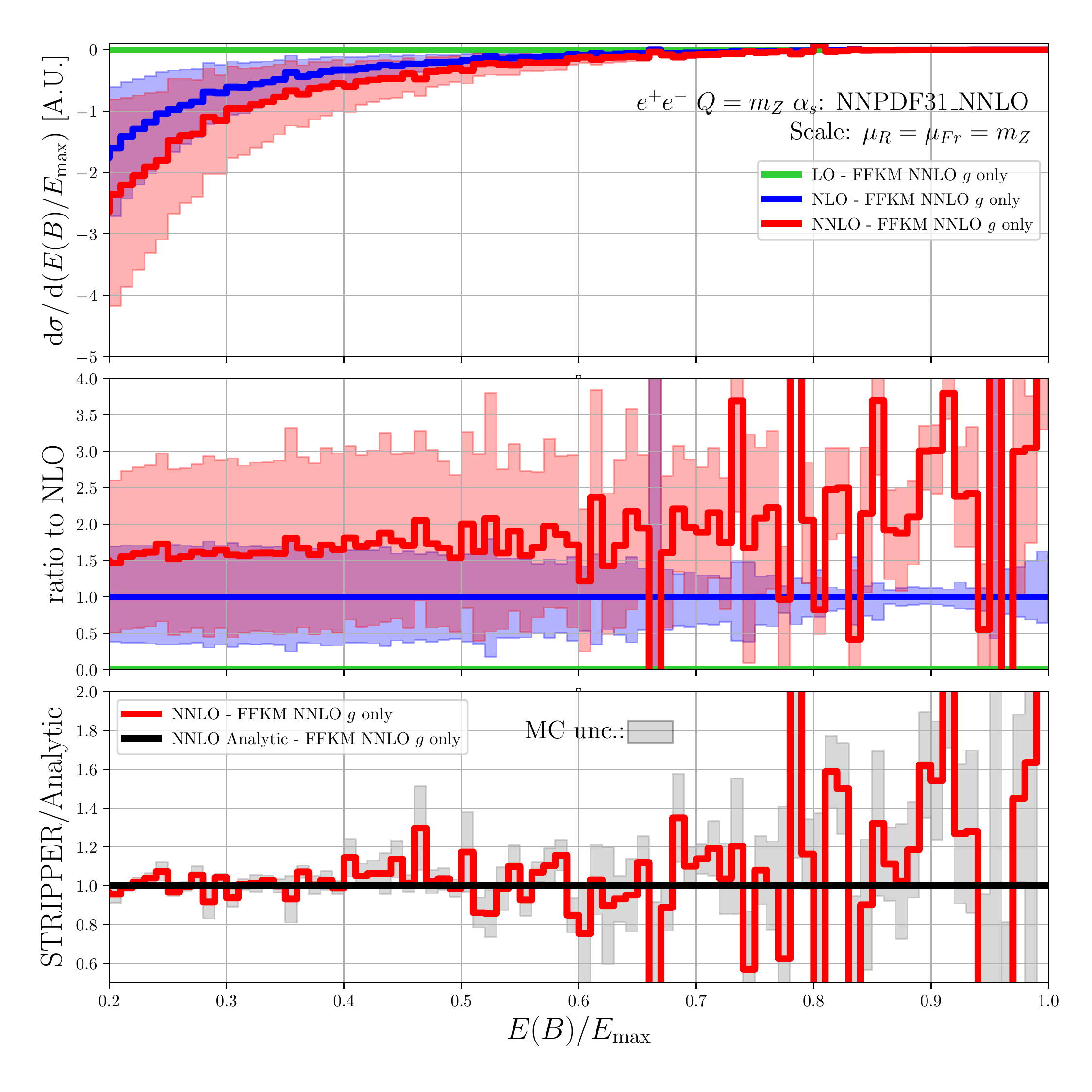}
  \caption{Comparison of predictions for the $B$-hadron energy spectrum in $e^+e^-$ collisions based on our numerical calculation and on the exact analytic $e^+e^-$ coefficient functions. A single partonic channel is shown at a time: $b+\bar b$ (left) and gluon (right).}
  \label{fig:e+e-}
\end{figure}

As an important check of our numerical setup we calculate the coefficient functions in $e^+e^-$ at NNLO QCD. In fig.~\ref{fig:e+e-} we show a typical $e^+e^-$ observable, the normalized $B$-hadron energy, computed at LO, NLO and NNLO QCD. It is compared at NNLO to a calculation of the same observable using the exact analytic form of the $e^+e^-$ coefficient functions \cite{Rijken:1996vr,Rijken:1996npa,Rijken:1996ns,Mitov:2006wy}. We check separately the quark and gluon coefficient functions by comparing the $B$-hadron energy including only the $b$ and $\bar b$ contributions (left) or only the gluon one (right). The $b+\bar b$ and $g$ contributions are separated according to eq.~(\ref{eq:factorized-x-section}).

The numerical setup is as follows. In all cases we use the FFKM fragmentation functions set at NNLO introduced in sec.~\ref{sec:PFF-NPFF}. The calculations are performed for a fixed central scale choice $\mu_R=\mu_{Fr}=m_Z$. The NNLO comparison between the two setups has been performed only for this central scale choice. The value of the strong coupling constant is taken from the {\tt LHAPDF} interface as supplied with the {\tt NNPDF3.1} pdf set. The numerical values of all other parameters entering the calculation are given in eq.~(\ref{eq:GF-scheme}). 

As is evident from fig.~\ref{fig:e+e-} there is an excellent agreement between the two calculations, within the MC error of the numeric calculation, and in the full kinematic range considered. This agreement represent a very strong check on the correctness of our numerical setup for both the quark and gluon coefficient functions.

\subsection{Sum rules in top decay}\label{sec:sum-rules}

It is well known \cite{Nason:1993xx} that heavy flavor production in $e^+e^-$ collisions satisfies the following sum rule:
\begin{equation}
\sigma={1\over 2}\sum_h\int_0^1dx\,x\,{d\sigma_h\over dx}\,,
\label{eq:sum-rule-e+e}
\end{equation}
where $h$ denotes a specific hadron that can be produced in the fragmentation of the heavy flavor and $x=2E(h)/E_{\rm had}$, with $E_{\rm had}=Q$ being the energy available for hadronic radiation. The fragmentation functions which are implicit in the above equation satisfy the following momentum conservation condition
\begin{equation}
\sum_h\int_0^1dz\,z\, D_{i\to h}(z) = 1\,.
\label{eq:sum-rule-D}
\end{equation}

As an additional check of our computational setup we verify that a sum rule analogous to eq.~(\ref{eq:sum-rule-e+e}) is satisfied in the case of $b$-production in top quark decay. To this end we construct a set of fake fragmentation functions that fulfill eq.~(\ref{eq:sum-rule-D}):
\begin{eqnarray}
D_b(z) &=& 60 z^2(1-z)^2\,,\nonumber\\
D_{\bar b}(z) &=& 105 z(1-z)^4\,,\nonumber\\
D_g(z) &=& 30 z(1-z)^2\,,\nonumber\\
D_q(z) &=& 168 z^4(1-z)^2\,,\nonumber\\
D_{\bar q}(z) &=& 504 z^4(1-z)^3\,.
\end{eqnarray}
For the purpose of checking the calculation of the coefficient functions it is sufficient to consider the case of a single hadron species. The equivalent of eq.~(\ref{eq:sum-rule-e+e}) for the case of top quark decay reads
\begin{equation}
\Gamma=\int_0^1dx\,x\,{d\Gamma\over dx}\,,
\label{eq:sum-rule-tdec}
\end{equation}
with $x=E(B)/E_{\rm had}$, where $E_{\rm had}=m_t-E_W$ is the energy available for hadronic radiation in top quark decay. The maximum value of $E(B)$ is given in eq.~(\ref{eq:EB-max}).

By comparing the RHS of eq.~(\ref{eq:sum-rule-tdec}) with an independent direct calculation of the top quark width we have verified that the pure NNLO correction of order ${\cal O}(\as^2)$ satisfies eq.~(\ref{eq:sum-rule-tdec}) with numerical precision of about 3\%.

\end{document}